\documentclass[pra,reprint,twocolumn,superscriptaddress,showpacs]{revtex4-1}

\usepackage{graphicx}
\usepackage{epsfig}
\usepackage{amsmath,bbm,amssymb,bm,times}
\usepackage[colorlinks,linkcolor=blue,citecolor=blue]{hyperref}

\newcommand{\eqn}[1]{
\begin{eqnarray}
	#1
\end{eqnarray}
}

\begin{document}
\title{Efficient variational approach to dynamics of a spatially extended  bosonic Kondo model}
\author{Yuto Ashida}
\email{ashida@ap.t.u-tokyo.ac.jp}
\affiliation{Department of Applied Physics, University of Tokyo, 7-3-1 Hongo, Bunkyo-ku, Tokyo 113-8656, Japan}
\affiliation{Department of Physics, University of Tokyo,  7-3-1 Hongo, Bunkyo-ku, Tokyo 113-0033, Japan}
\author{Tao Shi} 
\email{tshi@itp.ac.cn}
\affiliation{CAS Key Laboratory of Theoretical Physics, Chinese Academy of Sciences, Beijing 100190, China}
\author{Richard Schmidt}
\affiliation{Max-Planck-Institut f\"ur Quantenoptik, Hans-Kopfermann-Strasse. 1, 85748 Garching, Germany}
\affiliation{Munich Center for Quantum Science and Technology (MCQST), Schellingstr. 4, 80799 M\"unchen, Germany}
\author{H. R. Sadeghpour}
\affiliation{ITAMP, Harvard-Smithsonian Center for Astrophysics, Cambridge, MA 02138, USA}
\author{J. Ignacio Cirac}
\affiliation{Max-Planck-Institut f\"ur Quantenoptik, Hans-Kopfermann-Strasse. 1, 85748 Garching, Germany}
\affiliation{Munich Center for Quantum Science and Technology (MCQST), Schellingstr. 4, 80799 M\"unchen, Germany}
\author{Eugene Demler}
\affiliation{Department of Physics, Harvard University, Cambridge, MA 02138, USA}

\date{\today}

\begin{abstract} 
We develop an efficient  variational approach to studying dynamics of a localized quantum spin coupled to a bath of mobile spinful bosons. We use parity symmetry  to decouple the impurity spin from the environment via a canonical transformation and reduce the problem to a model of the interacting bosonic bath. We  describe coherent time evolution of the latter using bosonic Gaussian states as a variational ansatz. We provide full analytical expressions for equations describing variational time evolution that can be applied to study in- and out-of-equilibrium phenomena in a wide class of quantum impurity problems. In the accompanying paper [Y. Ashida {\it et al.}, Phys. Rev. Lett. 123, 183001 (2019)], we present a concrete application of this general formalism to the analysis of the Rydberg Central Spin Model, in which the spin-1/2 Rydberg impurity undergoes spin-changing collisions in a dense cloud of two-component ultracold bosons. To illustrate new features arising from orbital motion of the bath atoms, we compare our results to the Monte Carlo study of the model with spatially localized bosons in the bath, in which random positions of the atoms give rise to random couplings of the standard central spin model.
\end{abstract}

\maketitle

\section{Introduction}

Understanding dynamics of a quantum system coupled to a reservoir is a ubiquitous problem in modern physics. Models of this type can be used as a starting point for describing many important phenomena in condensed matter physics, atomic and molecular physics, and optics. Furthermore, coherent control and manipulation of quantum systems coupled to many-body reservoirs are the primary prerequisite for a successful implementation of quantum information processing. Current constraints on the applications of quantum technologies arise from practical limitations on coherence of quantum systems due to their interaction with external environments. 

In the context of many-body physics, a problem of particular importance is the model of quantum impurity interacting with a bosonic bath. This class of models can be used to describe a large number of physical systems, including electrons coupled to phonons  \cite{Pekar,FRP55,JTD_book}, spins coupled to a dissipative environment \cite{LAJ87}, the Bose Kondo problem \cite{FGM04,FS06,FFM11,FT15}, electrons in semiconductors \cite{Furdyna88,Lo1994,Vukmirovic2012}, and $^3$He-$^4$He mixtures \cite{BJ67}. Renewed interest in these problems
comes from recent experiments with ultracold atoms \cite{SA09,FB11,KoM12,Meinert945,Cetina2015,Cetina2016,SF16,HMG16,JNB16,YZZ19} that allow one to create many-body systems with tunable parameters and to obtain detailed characterizations of the dynamics. Theoretical studies lead to significant advances in understanding of polaron physics \cite{CFM06,RS12,EDK17,TJ09,NA10,CW11,CW112,RS13,VJ13,LW14,SYE162,Levinsen2015,Giorgini2015,Giorgini2016,CRS15,YA17,PLA18} beyond the conventional paradigm established in the previous studies of solid-state systems \cite{AAS_book,DJT09}. An important common feature of these systems is that bath modes are delocalized. Typical examples include phonons in crystal lattices or Bose-Einstein condensates (BEC). In the Bose Kondo model \cite{ZL04}, the bosonic spinful bath consists of paramagnons of the nearby antiferromagnetic phase and is thus delocalized.

Another important class of quantum baths features localized bath modes, such as nuclear spins that interact with the electron spin on a quantum dot  \cite{KA03,JS03,CWA04}. A paradigmatic model for describing such systems is the central spin model \cite{MG76,NVP00,TJM03,DJ04,EAY05,BM07,WZ07,CG07,LB08,BM10,WWM12,KEM12,FA13,RDA18}, which couples a central two-level system to localized (i.e., immobile) environmental spins. This  model is also commonly used to describe dynamics of spin qubits  realized with phosphorus impurities in silicon \cite{WWM10,TAM12} and nitrogen-vacancy centers in diamond \cite{HR08}. A particularly intriguing feature of the central spin model is its quantum integrability \cite{MG76}, which implies the existence of an extensive number of integrals of motion. Several powerful techniques have been developed to analyze dynamics of this model, including  Bethe ansatz \cite{MG76,LA01,BM07},  the Chebyshev-expansion method \cite{DVV03}, perturbative approaches \cite{CWA04,FJ07,FE08},  effective-Hamiltonian methods \cite{YW06,CL09},  cluster correlation expansion \cite{YW08}, the nonperturbative master-equation approach \cite{BE12} and a self-consistent Holstein-Primakoff approximation \cite{KEM12}. Relation between the central spin model and the BCS theory of superconductivity has also been actively explored \cite{Levitov04,EAY05}.

Until now the above two classes of many-body systems, namely, quantum impurity problems with mobile and localized bath modes have been analyzed separately. 
In particular, it is not obvious how to generalize the previous theoretical approaches developed in the central spin problem so as to handle the case of delocalized (i.e., mobile) environmental spins.
In this paper, we develop a variational approach to study a generalized type of a quantum impurity problem in which environmental bosons are mobile and spinful while interaction between the localized spin and the surrounding mobile spins is spatially extended.  A particularly intriguing aspect of this problem is the interplay between spin dynamics and orbital motion of host particles. A flip of the impurity spin dramatically changes the interaction between the impurity and environmental bosons, and vice versa, orbital motion of bosons affects spin-exchange processes because the interaction depends on the positions of atoms. 
 This unique interplay thus compounds the difficulties of including both orbital motion of mobile bosons and nonlocal spin interactions characteristic of the central spin model. To solve such a class of challenging problems, we describe a new variational approach that first decouples the impurity by employing the canonical transformation \cite{YA18L,*YA18B} (at the expense of introducing interactions among the bath particles) and then approximates coherent dynamics of the bath with the bosonic Gaussian states. %We recall that Gaussian states can include coherent expectation values of individual modes as well as all possible two-mode correlations. 
 We provide explicit analytical expressions for the nonlinear time-evolution equations of the vector of coherent expectation values and the covariance matrix of individual modes. 

The last two decades have witnessed remarkable developments in giant Rydberg impurities in ultracold atoms, offering a new experimental platform for quantum information processing \cite{Saffman10} and for realizing new types of strongly correlated many-body states \cite{GC00,ELH02,BV09,GA14,RS16,SM16,CF18,*RS18,KKS18}.
In the accompanying paper~\cite{YA19L}, we point out that Rydberg excitations in ultracold atoms naturally realize the new class of the quantum impurity problem, which we term as the Rydberg Central Spin Model (RCSM).  We here discuss in detail how our general theoretical approach can be applied to study nonequilibrium properties of the RCSM.
 While polaronic effects in Rydberg spectroscopy have recently attracted considerable attention, the focus of previous studies has been either on settings in which only a triplet scattering channel needs to be included and thus spin dynamics is frozen \cite{GC00,ELH02,BV09,GA14,RS16,SM16,CF18,*RS18,KKS18}, or on  low-density regimes, where only few-body molecular physics is relevant \cite{ADA14,SH15,BF16,MD19,EF19}.

 This paper is organized as follows. In Sec.~\ref{sec2}, we present details of the  variational approach to studying dynamics of a quantum spin coupled to a generic bosonic environment with the only requirement of having parity symmetry (e.g., symmetry of the rotation by $\pi$ around a certain axis). We introduce a canonical transformation that completely decouples the impurity from the environment. We present the formalism to approximate the time evolution of the environmental wavefunction in the transformed frame using a bosonic Gaussian state as a variational ansatz.  We derive analytical expressions of the equations of motion for the variational parameters. In Sec.~\ref{sec3}, we apply our general approach to analyze nonequilibrium dynamics  of the RCSM. We compute absorption spectrum that can be experimentally probed with Rydberg spectroscopy and  obtain real-time dynamics of the Rydberg-electron spin. To elucidate the crucial role of the orbital motion of bath atoms, we also analyze the central spin model with random couplings by employing the Monte Carlo sampling. Finally, we summarize the results and present an outlook in Sec.~\ref{sec4}.

\section{General formulation\label{sec2}}
\subsection{Impurity decoupling}
In this section, we formulate the variational approach in a general way so that it can be applied to a wide class of quantum impurity problems. Specifically, we consider the following class of Hamiltonians 
\eqn{\label{Hamiltonian}
\hat{H}=\sum_{n\alpha}\epsilon_{n\alpha}\hat{b}_{n\alpha}^{\dagger}\hat{b}_{n\alpha}+\hat{{\bf S}}_{e}\cdot\hat{{\bf S}}_{{\rm env}}+h_{z}\hat{S}_{e}^{z},
}
where $\hat{b}_{n\alpha}$ ($\hat{b}^\dagger_{n\alpha}$) are  bosonic annhiliation  (creation) operators of environmental modes $n=1,2,\ldots,N_b$ of energy $\epsilon_{n\alpha}$ with two internal degrees of freedom $\alpha=\Uparrow,\Downarrow$. We note that environmental modes are not specified here; they can be, for example, momentum eigenstates or single-particle bound states. The localized spin-1/2 operator is denoted by $\hat{{\bf S}}_{e}=\hat{\boldsymbol{\sigma}}_{e}/2$.   The second term represents a Kondo-type interaction between the localized spin and the environmental spins, in which we introduce the spin-density operator $\hat{\bf S}_{\rm env}$ including couplings as
\eqn{
\hat{S}_{{\rm env}}^{a}=\frac{1}{2}\sum_{mn\alpha\beta}g_{mn}^{a}\hat{b}_{m\alpha}^{\dagger}\sigma_{\alpha\beta}^{a}\hat{b}_{n\beta}
}
with $a=x,y,z$. Here, the $g_{mn}^a$ are arbitrary $N_b\times N_b$ Hermitian matrices labeled by $a$ and represent a generic form of an impurity-environment coupling that can in general be long-ranged and anisotropic. We note that this class of interaction reduces to the standard Kondo coupling~\cite{KJ64,HAC97,KM17} when it is local in space. The last term in Eq.~\eqref{Hamiltonian} is a magnetic-field term acting on the localized spin only. This term should be understood as a difference of Zeeman energies between the impurity and the bath spins. We note that the Hamiltonian commutes with the total spin $\hat{S}_{\rm tot}^z=\hat{S}_e^z+\hat{S}_{\rm env}^z$, and thus the term proportional to it merely shifts the spectrum by a global constant. In this paper, we set $h_z=0$ (except the results plotted in Fig.~\ref{fig3}). The magnetic term can be neglected if either the magnetic field is switched off or one uses atoms having the same Zeeman energies.

The first  key step in our variational approach is to employ the canonical transformation \cite{YA18L,*YA18B} to decouple the localized spin operator. 
Since the Hamiltonian~\eqref{Hamiltonian} satisfies the parity symmetry with respect to $\pi$ rotation of the entire system around $z$ axis, it commutes with the parity operator $\hat{{\rm P}}=\hat{\sigma}_{e}^{z}\hat{{\rm P}}_{{\rm env}}$ with $\hat{{\rm P}}_{{\rm env}}=e^{i\pi\hat{N}_{\Uparrow}}$ and $\hat{N}_{\Uparrow}=\sum_{n}\hat{b}_{n\Uparrow}^{\dagger}\hat{b}_{n\Uparrow}$. 
This conserved parity operator can be mapped onto the localized spin-1/2 operator via 
$\hat{U}^{\dagger}\hat{{\rm P}}\hat{U}=\hat{\sigma}_{e}^{x}$ with the unitary transformation~\cite{YA18L,*YA18B}:
\eqn{\label{unitary}
\hat{U}=\frac{1}{\sqrt{2}}(1+i\hat{\sigma}_{e}^{y}\hat{{\rm P}}_{{\rm env}}).} 
Thus, transforming to the `corotating' frame of the impurity via $\hat{U}$, the localized spin operator turns out to be a conserved quantity and its dynamics freezes, i.e., the system satisfies
\eqn{
[\hat{\tilde{H}},\hat{\sigma}_{e}^{x}]=0,
}
where $\hat{\tilde{H}}\equiv\hat{U}^{\dagger}\hat{H}\hat{U}$ is the transformed Hamiltonian.
Specifying the parity-symmetry sector ${\rm P}=\pm 1$, we can treat the spin operator as a classical number $\sigma_{e}^x=\pm1$ in the transformed frame. The resulting Hamiltonian thus only contains the environmental degrees of freedom:
\eqn{\label{transHamiltonian}
\hat{\tilde{H}}&=&\hat{\tilde{H}}_{0}+\hat{\tilde{H}}_{1},\\
\hat{\tilde{H}}_{0}&=&\sum_{n\alpha}\epsilon_{n\alpha}\hat{b}_{n\alpha}^{\dagger}\hat{b}_{n\alpha}+\frac{1}{4}\sigma_{e}^{x}\hat{S}_{{\rm env}}^{x},\label{H0}\\
\hat{\tilde{H}}_{1}&=&\frac{1}{2}\hat{{\rm P}}_{{\rm env}}\left(\sigma_{e}^{x}\hat{S}_{{\rm env}}^{z}-i\hat{S}_{{\rm env}}^{y}\right)+\frac{h_{z}}{2}\sigma_{e}^{x}\hat{{\rm P}}_{{\rm env}}.\label{H1}
}
Here, $\hat{\tilde{H}}_0$ represents the quadratic part of the transformed Hamiltonian while $\hat{\tilde{H}}_1$ represents interactions among environmental bosons that are in general multibody due to the nonlocality of the operator $\hat{\rm P}_{\rm env}$. The latter appears at the cost of the elimination of the impurity degree of freedom and can be interpreted as effective boson-boson interactions mediated via the impurity spin. 

\subsection{Variational principle\label{Sec_vari}}
The next step is to approximate the time-evolved bath state governed by the transformed Hamiltonian $\hat{\tilde{H}}$ by a tractable many-body wavefunction. To this end, we choose a bosonic Gaussian state $|\Psi_{\rm GS}\rangle$ \cite{WC12,ST17} as  an efficient variational state to describe the bath wavefunction in the transformed frame. 
The Gaussian states naturally include, as certain subclasses, variational states appropriate for the standard mean-field theory based on, e.g., the Gross-Pitaevskii equation and the Bogoliubov-de~Gennes equations. 

We here consider a generic Gaussian state, which is parameterized by a $4N_b$-dimensional real vector ${\boldsymbol \phi}$ and a $4N_b\times 4N_b$ real symmetric matrix $\Gamma$: 
\eqn{
\boldsymbol{\phi}&=&\langle\hat{\boldsymbol{\psi}}\rangle_{{\rm GS}},\\\Gamma&=&\frac{1}{2}\left\langle \left\{ \delta\hat{\boldsymbol{\psi}},\delta\hat{\boldsymbol{\psi}}^{{\rm T}}\right\} \right\rangle _{{\rm GS}},
}
where $\langle\cdots\rangle_{\rm GS}$ denotes the expectation value with respect to the Gaussian state (GS) and  $\hat{\boldsymbol{\psi}}=(\hat{\bm{x}},\hat{\bm{p}})^{{\rm T}}$ denotes a vector of the quadrature operators $\hat{x}_{n\alpha}=\hat{b}_{n\alpha}^{\dagger}+\hat{b}_{n\alpha}$ and $\hat{p}_{n\alpha}=i(\hat{b}_{n\alpha}^{\dagger}-\hat{b}_{n\alpha})$ as
\eqn{
\hat{\bm{x}}&=&\left(\hat{x}_{1\Uparrow},\cdots,\hat{x}_{N_{b}\Uparrow},\hat{x}_{1\Downarrow},\cdots,\hat{x}_{N_{b}\Downarrow}\right),\\
\hat{\bm{p}}&=&\left(\hat{p}_{1\Uparrow},\cdots,\hat{p}_{N_{b}\Uparrow},\hat{p}_{1\downarrow\Downarrow},\cdots,\hat{p}_{N_{b}\Downarrow}\right),
}
and  
$\delta\hat{\boldsymbol{\psi}}=\hat{\boldsymbol{\psi}}-\boldsymbol{\phi}$ is the fluctuation from the mean-field value. Physically, nonvanishing values of the vector $\boldsymbol \phi$ indicate macroscopic occupations of certain environmental modes as appropriate for a description of BEC. In addition, the covariance matrix $\Gamma$ in the Gaussian state describes the squeezing of environmental  bosons, enabling one to take into account quantum depletion arising from Bogoliubov excitations on top of a simple coherent state. Note that the number of parameters  grows at most quadratically with the number of environmental modes.

An explicit form of the Gaussian state is given by
\eqn{|\Psi_{{\rm GS}}\rangle=e^{i\theta}e^{\frac{i}{2}\hat{\boldsymbol{\psi}}^{{\rm T}}\sigma\boldsymbol{\phi}}e^{-\frac{i}{4}\hat{\boldsymbol{\psi}}^{{\rm T}}\xi\hat{\boldsymbol{\psi}}}|0\rangle\equiv\hat{U}_{{\rm GS}}|0\rangle,
\label{GSexp}
}
where $\sigma=i\sigma^{y}\otimes{\rm I}_{2N_{b}}$ with ${\rm I}_d$ being the $d\times d$ unit matrix ($d$ is a positive integer) and  $\xi$ is a $4N_b\times 4N_b$ real-symmetric matrix. A matrix dimension is $4N_b$ ($2N_b$) if a matrix acts on spinful (spinless) single-particle modes. These matrices can be related to the covariance matrix via
\eqn{\Gamma&=&\gamma\gamma^{{\rm T}},\\
\gamma&=&e^{\sigma\xi}.
}
Note that in Eq.~\eqref{GSexp} we explicitly include a phase factor $\theta$, which is necessary to obtain the absorption spectrum of the system as detailed later. 

The time-evolution equation can be obtained from the time-dependent variational principle \cite{JR79,KP08,ST17}. Specifically, we project the  exact real-time evolution of the environmental state (in the transformed frame),
\eqn{
i\hbar\partial_{t}|\Psi(t)\rangle=\hat{\tilde{H}}|\Psi(t)\rangle,
}
on the manifold spanned by the present variational states. This procedure is equivalent to minimizing the deviation $\epsilon=\Vert (i\hbar\partial_{t}-\hat{\tilde{H}})|\Psi_{{\rm GS}}(t)\rangle\Vert^{2}$ from the exact evolution at each moment, leading to
\eqn{i\hbar\partial_{t}|\Psi_{{\rm GS}}(t)\rangle=\hat{P}_{{\rm \partial}}\hat{\tilde{H}}|\Psi_{{\rm GS}}(t)\rangle,\label{realtime}
}
where  $\hat{P}_\partial$ is a projection operator onto the tangent space of the variational manifold. 
The time derivative of the Gaussian state can be rewritten as
\eqn{\label{GSderiv}
\partial_{t}|\Psi_{{\rm GS}}(t)\rangle=\hat{U}_{{\rm GS}}&\Bigl(&c+\frac{i}{2}\hat{\boldsymbol{\psi}}^{{\rm T}}\gamma^{{\rm T}}\sigma\partial_{t}\boldsymbol{\phi}
\nonumber\\
&+&\frac{i}{4}:\hat{\boldsymbol{\psi}}^{{\rm T}}\gamma^{{\rm T}}\sigma(\partial_{t}\gamma)\hat{\boldsymbol{\psi}}:\Bigr)|0\rangle
}
with  $c=id_{t}\theta+(i/4){\rm Tr}[\gamma^{{\rm T}}\sigma\left(\partial_{t}\gamma\right)\Gamma]$ being a scalar while the state acted by the transformed Hamiltonian can be represented as
\eqn{\label{HtildeGS}
\hat{\tilde{H}}|\Psi_{{\rm GS}}(t)\rangle=\hat{U}_{{\rm GS}}&\Bigl(&\langle\hat{\tilde{H}}\rangle_{{\rm GS}}+\frac{1}{2}\hat{\boldsymbol{\psi}}^{{\rm T}}\gamma^{{\rm T}}{\cal H}_{\phi}
\nonumber\\
&+&\frac{1}{4}:\hat{\boldsymbol{\psi}}^{{\rm T}}\gamma^{{\rm T}}{\cal H}_{\Gamma}\gamma\hat{\boldsymbol{\psi}}:+\delta\hat{O}\Bigr)|0\rangle.
}
Here, $:\;:$ indicates the normal order of the bosonic operators $\hat{b}_{n\alpha}$ and $\hat{b}^\dagger_{n\alpha}$ and we introduce the functional derivatives of the variational energy as
\eqn{
{\cal H}_{\phi}&\equiv&2\frac{\delta\langle\hat{\tilde{H}}\rangle_{{\rm GS}}}{\delta\boldsymbol{\phi}\label{Hphi}},\\
{\cal H}_{\Gamma}&\equiv&4\frac{\delta\langle\hat{\tilde{H}}\rangle_{{\rm GS}}}{\delta\Gamma}\label{HGamma}.
}
The operator $\hat{P}_\partial$ projects out the  higher-order contributions of $\hat{\boldsymbol \psi}$, which are denoted as $\delta\hat{O}$ in Eq.~\eqref{HtildeGS}. Using Eqs.~\eqref{GSderiv} and \eqref{HtildeGS} and  
comparing linear and quadratic terms in $\hat{\boldsymbol \psi}$ on both sides of Eq.~\eqref{realtime},
 we obtain the  set of variational equations:
\eqn{\hbar d_{t}\boldsymbol{\phi}&=&\sigma{\cal H}_{\phi},\label{eqphi}\\
\hbar d_{t}\Gamma&=&\sigma{\cal H}_{\Gamma}\Gamma-\Gamma{\cal H}_{\Gamma}\sigma.\label{eqGamma}}
We emphasize that these variational equations should in general offer better results than the standard mean-field theories since the present approach is based on a general Gaussian state and thus can autonomously take into account all the possible two-point correlations on top of an arbitrary BEC state in an unbiased manner. 

\subsection{Equations of motion}
To integrate the real-time evolutions~\eqref{eqphi} and \eqref{eqGamma} of the time-dependent variational parameters $\boldsymbol \phi$ and $\Gamma$, we need analytical expressions of the functional derivatives ${\cal H}_{\phi}$ and ${\cal H}_{\Gamma}$. 
To this end, we first express the expectation values of the quadratic  and  interaction Hamiltonians in the transformed frame in terms of the variational parameters as follows:
\eqn{
\langle\hat{\tilde{H}}_{0}\rangle_{{\rm GS}}&=&\frac{1}{4}{\rm Tr}\left[{\cal H}_{0}\Gamma\right]+\frac{1}{4}\boldsymbol{\phi}^{{\rm T}}{\cal H}_{0}\boldsymbol{\phi}-\frac{1}{4}{\rm Tr}\left[{\cal H}_{0}\right],\\
\langle\hat{\tilde{H}}_{1}\rangle_{{\rm GS}}&=&\frac{1}{4}{\rm Tr}\left[\Sigma_{g}^{{\rm T}}\Omega\right]+\frac{h_{z}}{2}\sigma_{e}^{x}\langle\hat{{\rm P}}_{{\rm env}}\rangle_{{\rm GS}},\label{Htilde1GS}
}
where we introduce a $4N_b\times 4N_b$ real-symmetric matrix ${\cal H}_{0}={\cal S}[({\rm I}_{2}-\sigma^y)\otimes \tilde{h}_{0}]$ with ${\cal S}[A]\equiv(A+A^{{\rm T}})/2$ being the matrix symmetrization and $\tilde{h}_0$ being the single-particle Hamiltonian in the transformed frame (cf. Eq.~\eqref{H0}):
\eqn{
\tilde{h}_{0}={\rm diag}(\epsilon_{n\alpha})+(\sigma_{e}^{x}/4)\sigma^{x}\otimes g_{mn}^{x}.
}
In the interaction energy~\eqref{Htilde1GS}, we introduce a $2N_b\times 2
N_b$ matrix 
\eqn{\Sigma_{g}=\sigma_{e}^{x}\sigma^{z}\otimes g_{mn}^{z}-i\sigma^{y}\otimes g_{mn}^{y}}
and define the $2N_b\times 2N_b$ matrix $\Omega$ including the environmental  parity operator by \cite{ST17}
\eqn{\Omega&\equiv&\langle\hat{{\rm P}}_{{\rm env}}\hat{{\bm b}}^{\dagger}\hat{{\bm b}}\rangle_{{\rm GS}}\nonumber\\
&=&-\Sigma_{z}\langle\hat{{\rm P}}_{{\rm env}}\rangle_{{\rm GS}}({\rm I}_{2N_{b}},-i{\rm I}_{2N_{b}})\nonumber\\
&\times&\!\left(\Gamma_{B}^{-1}\right)^{{\rm T}}\!\left[\frac{1}{2}(\Gamma-{\rm I}_{4N_{b}})\!+\!\boldsymbol{\phi}\boldsymbol{\phi}^{{\rm T}}\Gamma_{B}^{-1}\right]\left(\begin{array}{c}
{\rm I}_{2N_{b}}\\
i{\rm I}_{2N_{b}}
\end{array}\right).
}
Here, we represent the vector of the bosonic creation operators of different environmental modes as
\eqn{
\hat{{\bm b}}&=&(\hat{b}_{1\Uparrow},\cdots,\hat{b}_{N_{b}\Uparrow},\hat{b}_{1\Downarrow},\cdots,\hat{b}_{N_{b}\Downarrow}),
}
and introduce matrices $\Sigma_{z}=\sigma^{z}\otimes{\rm I}_{N_{b}}$, $\Lambda={\rm I}_{2}\otimes\Sigma_{z}$ and $\Gamma_{B}=({\rm I}_{4N_{b}}+\Lambda)\Gamma+{\rm I}_{4N_{b}}-\Lambda$.
The expectation value of the environmental parity operator is given by
\eqn{\langle\hat{{\rm P}}_{{\rm env}}\rangle_{{\rm GS}}=\frac{1}{\sqrt{{\rm det}(\Gamma_{B}/2)}}e^{-\frac{1}{2}\boldsymbol{\phi}^{{\rm T}}\Gamma_{B}^{-1}(1+\Lambda)\boldsymbol{\phi}}.\label{Penv}
}
Calculating the derivatives of the variational energy $\langle\hat{\tilde{H}}\rangle_{\rm GS}$ with respect to the parameters $\boldsymbol \phi$ and $\Gamma$ in Eqs.~\eqref{Hphi} and \eqref{HGamma}, 
 we now obtain the analytical expressions of ${\cal H}_\phi$ and ${\cal H}_\Gamma$ as follows:
\begin{widetext}
\eqn{{\cal H}_{\phi}&=&{\cal H}_{0}\boldsymbol{\phi}-2\langle\hat{\tilde{H}}_{1}\rangle_{{\rm GS}}\Gamma_{B}^{-1}(1+\Lambda)\boldsymbol{\phi}-\langle\hat{{\rm P}}_{{\rm env}}\rangle_{{\rm GS}}\Gamma_{B}^{-1}{\cal S}\left[{\cal G}\right](\Gamma_{B}^{-1})^{{\rm T}}\boldsymbol{\phi},\label{Hphia}\\
{\cal H}_{\Gamma}&=&{\cal H}_{0}+{\cal S}\left[2\langle\hat{\tilde{H}}_{1}\rangle_{{\rm GS}}\Gamma_{B}^{-1}(1+\Lambda)\left(\Phi-{\rm I}_{4N_{b}}\right)+\langle\hat{{\rm P}}_{{\rm env}}\rangle_{{\rm GS}}\Gamma_{B}^{-1}{\cal G}(\Gamma_{B}^{-1})^{{\rm T}}\left(2\Phi-{\rm I}_{4N_{b}}\right)\right],\label{HGammaa}
}
\end{widetext}
where we introduce the $4N_b\times 4N_b$ matrices 
\eqn{\Phi&=&\boldsymbol{\phi}\boldsymbol{\phi}^{{\rm T}}\Gamma_{B}^{-1}(1+\Lambda),\\
{\cal G}&=&\left(\begin{array}{c}
{\rm I}_{2N_{b}}\\
-i{\rm I}_{2N_{b}}
\end{array}\right)\Sigma_{z}\Sigma_{g}({\rm I}_{2N_{b}},i{\rm I}_{2N_{b}}).
}
We note that the variational equations~\eqref{eqphi} and \eqref{eqGamma} together with the  analytical formulae~\eqref{Hphia} and \eqref{HGammaa} are general and can be readily applied to studying out-of-equilibrium dynamics of a quantum system coupled to various types of environments. While we focus on the real-time evolution in this paper, the ground-state properties can also be analyzed by using Eqs.~\eqref{Hphia} and \eqref{HGammaa} to integrate the variational imaginary-time evolution whose explicit form is given in App.~\ref{appimag}.

\subsection{Absorption spectrum}
Out-of-equilibrium properties can be studied by analyzing the absorption spectrum \cite{Schmidt_2018}
\eqn{A(\omega)={\rm Re}\left[\int_{0}^{\infty}dte^{i\omega t}S(t)\right],\label{absorption}}
where $S(t)$ is the overlap between the initial state and a time-evolved state
\eqn{S(t)=\langle\Psi(0)|e^{-i\hat{\tilde{H}}t/\hbar}|\Psi(0)\rangle.}
For the sake of simplicity, a contribution from the free time evolution (e.g., the one without Rydberg interactions in the model discussed later) is not included here as it just shifts $A(\omega)$ by a  trivial constant.
To calculate $S(t)$ in the present variational approach, we have to obtain an equation of motion of the phase factor $\theta$ of the Gaussian environmental state in addition to that of the mean-field vector $\boldsymbol \phi$ and the covariance matrix $\Gamma$.  To  this end, the most convenient way is to parameterize the Gaussian state in the basis of $\hat{b}_{n\alpha}$ and $\hat{b}^\dagger_{n\alpha}$ operators as follows:
\eqn{|\Psi_{{\rm GS}}\rangle=e^{i\theta}e^{\frac{i}{2}\hat{\boldsymbol{\psi}}^{{\rm T}}\sigma\boldsymbol{\phi}}e^{\hat{{\bm b}}^{\dagger{\rm T}}\Xi\hat{{\bm b}}^{\dagger}}e^{\hat{{\bm b}}\Xi'\hat{{\bm b}}^{\dagger}}e^{\hat{{\bm b}}\Xi''\hat{{\bm b}}^{{\rm T}}}|0\rangle,
}
where $\Xi$, $\Xi'$ and $\Xi''$ are $2N_b\times2N_b$ matrices. The overlap can then be  obtained from
\eqn{S(t)=e^{i\theta}e^{-\frac{1}{2}|\delta\boldsymbol{\beta}|^{2}}e^{\delta\boldsymbol{\beta}^{\dagger}\Xi\delta\boldsymbol{\beta}^{*}},\label{st}
}
where $\delta{\boldsymbol \beta}$ is the difference mean-field vector in the basis of $(\hat{\bm b},\hat{\bm b}^\dagger)^{\rm T}$ defined as follows:
\eqn{\delta\boldsymbol{\beta}(t)&=&\boldsymbol{\beta}(t)-\boldsymbol{\beta}(0),\\
\boldsymbol{\beta}&=&\frac{1}{2}B^{\dagger}\boldsymbol{\phi}\label{beta}
}
with $B$ being the $4N_b\times 4N_b$ matrix 
\eqn{B=\left(\begin{array}{cc}
{\rm I}_{2N_{b}} & {\rm I}_{2N_{b}}\\
-i{\rm I}_{2N_{b}} & i{\rm I}_{2N_{b}}
\end{array}\right).}
The time evolution of the variational parameters $\delta {\boldsymbol \beta}$  is obtained by integrating the equations of motion~\eqref{eqphi} to yield ${\boldsymbol \phi}(t)$ and by applying Eq.~\eqref{beta} to transform it into an expression in the basis of $\hat{b}_{n\alpha}$ and $\hat{b}^\dagger_{n\alpha}$ operators. Following the same procedures relying on the time-dependent variational principle as discussed in Sec.~\ref{Sec_vari}, we  obtain the set of time-evolution equations for the parameters $\theta$ and $\Xi$,
\eqn{\hbar d_{t}\theta&=&-\langle\hat{\tilde{H}}\rangle_{{\rm GS}}+\frac{1}{4}\delta\boldsymbol{\phi}^{{\rm T}}{\cal H}_{\phi}+\frac{1}{4}{\rm Tr}\left[{\cal H}_{\Gamma}\Gamma\right]\nonumber\\
&&-\frac{1}{2}{\rm Tr}[h_{\Gamma}]-{\rm Tr}\left[h'^{*}_{\Gamma}\Xi\right],\label{eqtheta}\\
i\hbar d_{t}\Xi&=&\frac{1}{2}h_{\Gamma}'+h_{\Gamma}\Xi+\Xi h_{\Gamma}+2\Xi h'^{*}_{\Gamma}\Xi.\label{eqXi}
}
Here we denote 
$\delta\boldsymbol{\phi}(t)={\boldsymbol \phi}(t)-{\boldsymbol \phi}(0)$ and define 
the $2N_b\times 2N_b$ matrices $h_{\Gamma}$ and $h'_{\Gamma}$ by
\eqn{\left(\begin{array}{cc}
h_{\Gamma} & h'_{\Gamma}\\
h'^{*}_{\Gamma} & h_{\Gamma}
\end{array}\right)=\frac{1}{2}B^{\dagger}{\cal H}_{\Gamma}B.\label{hG}}
Integrating Eqs.~\eqref{eqtheta} and \eqref{eqXi} together with Eqs.~\eqref{Hphia} and \eqref{HGammaa}, we  calculate the overlap $S(t)$ from Eq.~\eqref{st} whose Fourier transform provides the absorption spectrum~\eqref{absorption}.
\section{Application to the Rydberg Central Spin Model\label{sec3}}
\begin{figure*}[t]
\includegraphics[width=150mm]{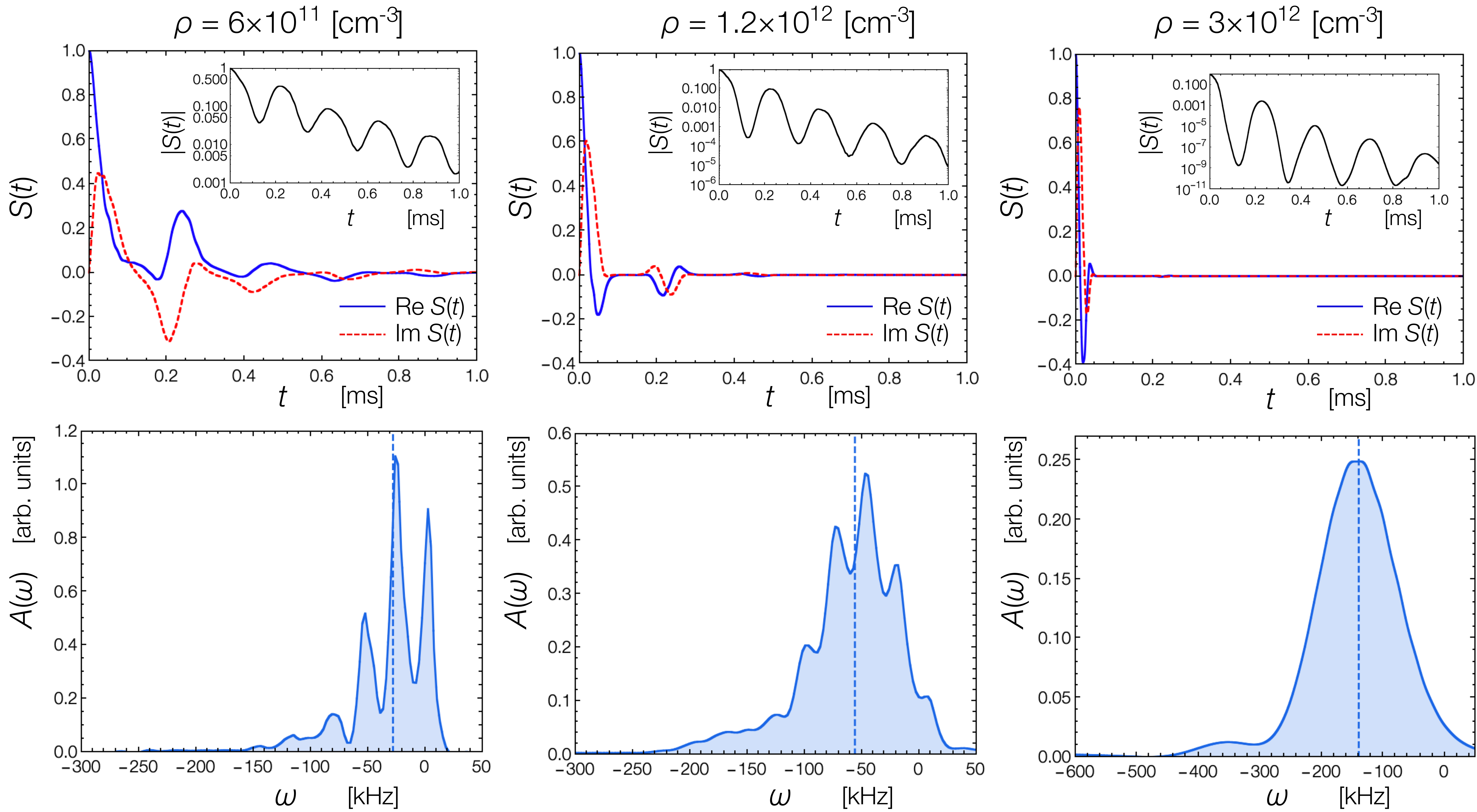} 
\caption{\label{fig1}
Applications of the  variational approach to nonequilibrium dynamics of a Rydberg impurity interacting with surrounding bosons via the anisotropic central  spin coupling. The obtained overlaps $S(t)$ (top panels) and the corresponding absorption spectra $A(\omega)$ (bottom panels) are plotted at different densities. In the top panels, the blue solid curves (red dashed curves) show the real (imaginary) parts of the overlaps. The absolute values $|S(t)|$ are shown in the insets. In the bottom panels, the dashed lines indicate the mean-field (MF) shifts $\Delta_{\rm MF}$ (cf. Eqs.~\eqref{meanF} and \eqref{meanFH}). 
}
\end{figure*}
\subsection{Variational results with orbital motion}
We now discuss the application of our general variational approach to a new class of the quantum impurity problem, in which bath bosons are mobile and spinful while interaction between the localized spin and the surrounding particles is spatially extended. This class of many-body problems can be naturally  realized with Rydberg gases~\cite{YA19L}. 
Rydberg excitations have been powerful probes for studying new regimes of out-of-equilibrium dynamics of molecular states and quantum many-body systems \cite{GC00,ELH02,BV09,GA14,RS16,SM16,CF18,*RS18,KKS18,ADA14,SH15,BF16,MD19,EF19}.
When an atom in a BEC is excited into a Rydberg state with a high principal quantum number $n$, the surrounding environmental bosons are subject to Fermi's pseudopotential \cite{EF34} created by frequent elastic scattering of the weakly bound Rydberg electron. If the electron-atom scattering length is negative, the interaction between the Rydberg electron and the environmental atoms is attractive and the atoms can become bound to the Rydberg excitations. This can lead to giant molecular states known as Rydberg molecules whose size can be several thousand of the Bohr radius \cite{GC00,BV09}. 

We consider a situation in which internal dynamics of environmental bosons is restricted to two internal states $\alpha=\Uparrow,\Downarrow$. Thus, the bosons act as  mobile pseudospin-1/2 particles and their scattering with the Rydberg electron separates into two channels.  We denote these scattering channels  as $1$ and $2$, and the corresponding pseudopotentials as $V_{1,2}$. The resulting Hamiltonian is (cf. Ref.~\cite{YA19L})
\eqn{\label{Rydberg}
\hat{H}\!=\!  \sum_{\alpha=\Uparrow,\Downarrow}\int d{\bf r}\hat{\Psi}_{{\bf r}\alpha}^{\dagger}{h}_{0}\hat{\Psi}_{{\bf r}\alpha}\!+\!\!\sum_{a=x,y,z}\hat{S}_{e}^{a}\int d{\bf r}g_{{\bf r}}^{a}\hat{S}_{{\bf r}}^{a}\!+\!h_{z}\hat{S}_{e}^{z},\nonumber\\
}
where $\hat{\Psi}_{{\bf r}\alpha}$ ($\hat{\Psi}_{{\bf r}\alpha}^\dagger$) are annihilation (creation) operators of bosons at position $\bf r$ with internal state $\alpha=\Uparrow,\Downarrow$, $\hat{\bf S}_e=\hat{\boldsymbol \sigma}_e/2$ is the spin operator of the Rydberg electron, $h_z$ is a magnetic field acting on the impurity spin, and  $\hat{\bf S}_{\bf r}$ is the environmental spin density
\eqn{
\hat{{\bf S}}_{{\bf r}}&=&\sum_{\alpha,\beta=\Uparrow,\Downarrow}\hat{\Psi}_{{\bf r}\alpha}^{\dagger}\left(\frac{\boldsymbol{\sigma}}{2}\right)_{\alpha\beta}\hat{\Psi}_{{\bf r}\beta}.
}
We introduce ${h}_0$ as the matrix elements of a single-particle Hamiltonian  
\eqn{
{h}_{0}&=&-\frac{\hbar^{2}\nabla^{2}}{2m}+V_{0}({\bf r}),
}
where $m$ denotes the atomic mass.
The potential $V_0$ and the long-range Kondo coupling $g_{\bf r}^a$ are in general given by linear combinations of the two Rydberg potentials $V_{1,2}$, in which  the coefficients depend on the internal structure of environmental atoms. As appropriate for the simplest setup of the scattering between two spin-1/2 particles, hereafter we choose  
$V_{0}=(3V_{1}+V_{2})/4$ and $g_{{\bf r}}^{z}= g_{\bf r}^\parallel\equiv V_{1}-V_{2}$ with $V_1$ and $V_2$ being the triplet- and singlet-pseudopotentials characterizing the interaction between the Rydberg electron and the surrounding atoms. As a representative case, we use the data for the potential profiles of the scattering between the  electron of the $^{87}$Rb$(87s)$ state and the ground-state $^{87}$Rb atoms; all the numerical results presented below are obtained for the triplet and singlet Rydberg potentials of Rb atoms with $n=87$. Yet, in practice we envision using two-electron environmental atoms (e.g., Sr atoms), in which the hyperfine coupling can be neglected as discussed in the accompanying paper~\cite{YA19L}. The potential profiles in this two-electron atomic setup are analogous to those in the Rb setup used here \cite{Priv}. Thus, essential features of the RCSM discussed below, which are qualitatively insensitive to these specific choices of potentials, will remain in the two-electron atomic setup. Meanwhile, a possible difference in using the two-electron atom is the anisotropy in the spin interaction~\cite{YA19L}. To address this effect, we set 
$g_{{\bf r}}^{x,y}= g_{\bf r}^{\perp}\equiv\sqrt{2}g_{{\bf r}}^{z}$. We neglect a temperature effect, boson-boson interactions and the ion recoil because  the Rydberg potentials typically exceed those energy scales. We also omit contributions from  $p$-wave scattering. 
%We remark that the  aim of the present work is to reveal essential features of the RCSM, which are qualitatively insensitive to a specific choice of the Rydberg state as long as the above conditions are satisfied.

Previous works on Rydberg molecule and polaron formation in a high-density regime \cite{GC00,ELH02,BV09,GA14,RS16,SM16,CF18,*RS18,KKS18} have exclusively focused on the following polarized initial state
\eqn{\label{triplet}
|\Psi_{\rm triplet}\rangle=|\!\uparrow\rangle_{e}|{\rm BEC}_{\Uparrow}\rangle,
}
where only the triplet channel is relevant to the dynamics. Thus, the interaction part of Eq.~\eqref{Rydberg} plays no role and the resulting Hamiltonian is quadratic. In this case, the spin dynamics is completely frozen and the orbital motion of environmental atoms solely characterizes the nonequilibrium properties of the system.

In the present work, we are interested in a different initial state with the Rydberg spin and the surrounding bosons being polarized into the opposite directions:
\eqn{\label{initial}
|\Psi_{0}\rangle=|\uparrow\rangle_{e}|{\rm BEC}_{\Downarrow}\rangle.
}
We note that such an initial condition has already been realized in the experiments at low-density regimes \cite{ADA14,SH15,BF16,MD19,EF19}.
In this setting, the central  spin-type interaction  (cf. the second term on the right-hand side of Eq.~\eqref{Rydberg}) now plays an important role and the nonequilibrium properties of the system are characterized by the interplay between the spin and orbital degrees of freedom of the environmental particles. 

The main difficulty here is taking into account mesoscopic collective response of the orbital motion as well as the  entanglement of the impurity and environmental spins. The main idea of the variational approach presented in this paper is to  employ the disentangling transformation~\eqref{unitary} to decouple the impurity spin and reduce the problem to that of the bath degrees of freedom with additional interactions. We then approximate the evolution of the environmental wavefunction in the transformed frame by the Gaussian state, which guarantees that all two-particle correlations are taken into account.

To apply the general formulation presented in the previous section, it is useful to introduce single-particle energy eigenstates as the computational basis:
\eqn{
{h}_{0}\psi_{n{\bf r}}&=&\epsilon_{n}\psi_{n{\bf r}},\\
\hat{b}_{n\alpha}^{\dagger}&=&\int d{\bf r}\psi_{n{\bf r}}\hat{\Psi}_{{\bf r}\alpha}^{\dagger},
}
where $n=1,2,\ldots,N_{b}$ is the nodal quantum number of the single-particle state  $\psi_{n{\bf r}}$. While we focus on the situation in which only states with zero angular momentum $l=0$ are relevant, the formulation can be easily generalized to include effects of nonzero angular momentum. Further details about the calculations of the eigenenergies $\epsilon_n$ and the wavefunctions $\psi_{n{\bf r}}$ are given in App.~\ref{appryd}. The central  spin couplings in this basis are given by
\eqn{g_{mn}^{a}=\int d{\bf r}g_{{\bf r}}^{a}\psi_{m{\bf r}}^{*}\psi_{n{\bf r}}}
with $a=x,y,z$. 
These identifications lead to the Hamiltonian~\eqref{Hamiltonian},  and Eqs.~\eqref{transHamiltonian}, \eqref{H0} and \eqref{H1} after performing the decoupling transformation~\eqref{unitary}.
The initial state~\eqref{initial} corresponds to the following initial values of the variational parameters:
\eqn{\label{initialphi}\boldsymbol{\phi}(t=0)&=&\sqrt{N}B\left(\begin{array}{c}
\int d{\bf r}\psi_{n{\bf r}}^{*}\psi_{{\rm ini},{\bf r}}\delta_{\alpha\Downarrow}\\
\int d{\bf r}\psi_{n{\bf r}}\psi_{{\rm ini},{\bf r}}^{*}\delta_{\alpha\Downarrow}
\end{array}\right),\\
\Gamma(t=0)&=&{\rm I}_{4N_{b}},
}
where $N$ is the total number of environmental bosons, which can be related to the density $\rho$, and $\psi_{{\rm ini},{\bf r}}$ is the initial lowest-energy wavefunction without Rydberg potentials (see App.~\ref{appryd} for further details). We note that this value represents the number of bosons in the box used to calculate eigenstates rather than the number of bosons in the Rydberg potential. Because the initial state~\eqref{initial} resides in the parity sector $\rm P=+1$, we can treat the Rydberg spin as a classical number $\sigma_{e}^x=+1$ in the transformed frame.

 \begin{figure}[b]
\includegraphics[width=56mm]{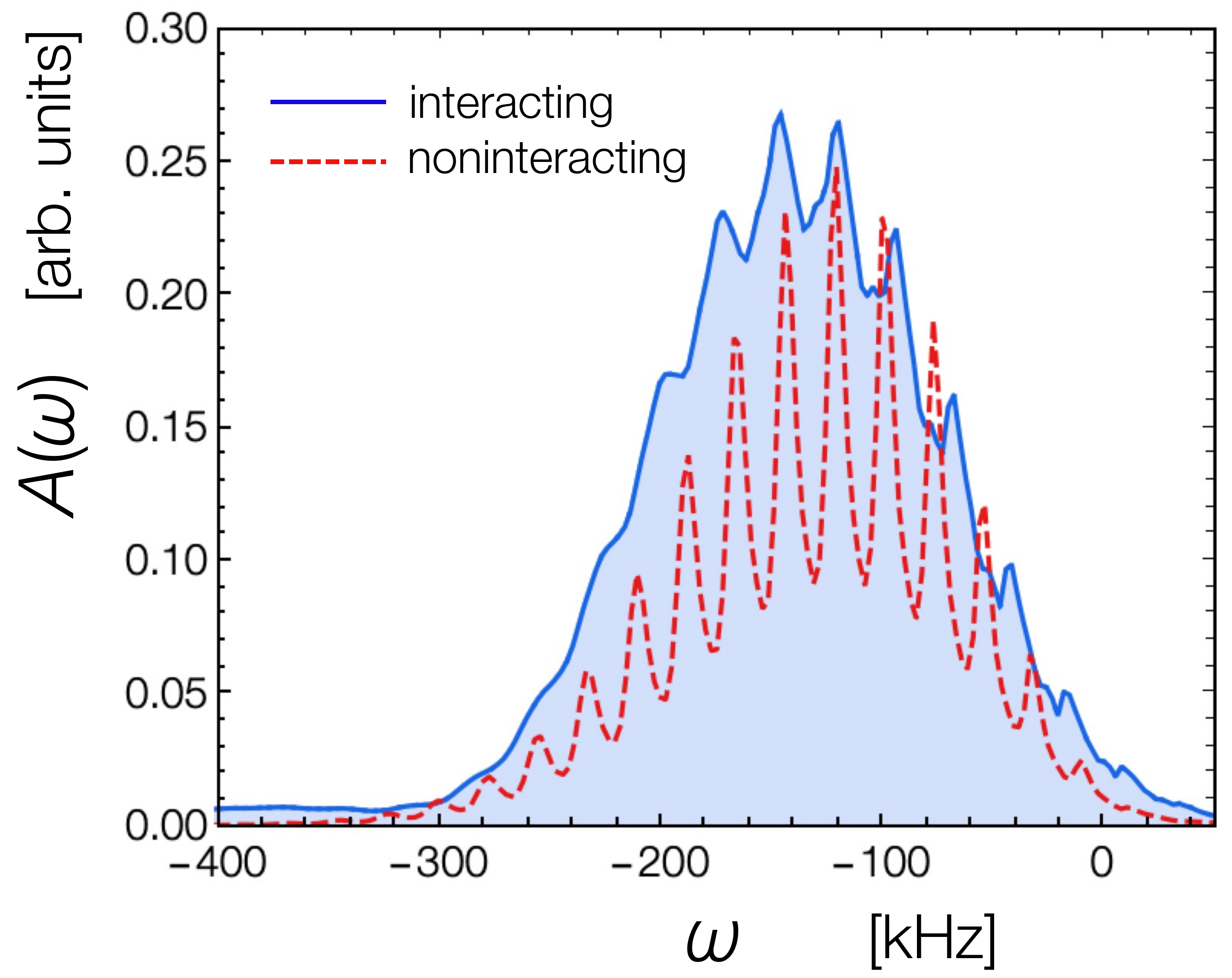} 
\caption{\label{fig_rev}
The comparison between the absorption spectra obtained from the variational approach for the full interacting Hamiltonian $\hat{H}$ (blue) and the results for the noninteracting (quadratic) Hamiltonian $\hat{H}_{\parallel}$ (red). The density is $\rho=3\times 10^{12}$cm${^{-3}}$. We use the isotropic interaction as consistent with Fig.~3 in the accompanying paper~\cite{YA19L} and rescale the overall factors of $A(\omega)$ for the sake of comparison between the two cases.
}
\end{figure}

To obtain values of the time-dependent variational parameters,
we integrate the variational time-evolution equations~\eqref{eqphi}, \eqref{eqGamma}, \eqref{eqtheta} and \eqref{eqXi} with the use of the analytical expressions of the functional derivatives~\eqref{Hphia} and \eqref{HGammaa}. We then calculate the overlap $S(t)$ from Eq.~\eqref{st} and obtain its Fourier spectrum $A(\omega)$ from Eq.~\eqref{absorption}. The magnetization dynamics of the central spin 
\eqn{m_z(t)\equiv\langle\hat{\sigma}_e^z(t)\rangle=\sigma_e^x\langle\hat{\rm P}_{\rm env}\rangle_{\rm GS}}
can be obtained from Eq.~\eqref{Penv}. To improve the numerical accuracy, we find it useful to implement a penalty term to ensure the spin conservation (see App.~\ref{apppenalty} for details). A typical number $N_b$ of single-particle states necessary to converge the variational results is an order of several tens of states; for instance, $N_b\sim80$ is sufficient for the parameters considered here.

Figure~\ref{fig1} shows the results obtained for the overlap $S(t)$ and the corresponding absorption spectrum $A(\omega)$ at different densities $\rho$.
As shown in the top panels in Fig.~\ref{fig1}, we find a fast decay in $S(t)$ that signals the rapid dephasing dynamics due to the creation of high-energy excitations of the environmental bosons. 
The fast decay of $S(t)$ in the time domain corresponds to the emergence of a Gaussian profile of the Fourier spectra $A(\omega)$ in the frequency domain in the   high-density regime  as shown in the bottom panels in Fig.~\ref{fig1}. As indicated by the dashed lines in the same panels, the centers of the Gaussian profiles agree with the following mean-field shift which neglects the flip-flop coupling:
\eqn{
\Delta_{\rm MF}&=&\langle\Psi_0|\hat{H}_\parallel|\Psi_0\rangle,\label{meanF}\\
\hat{H}_\parallel&=&\sum_{\alpha=\Uparrow,\Downarrow}\int d{\bf r}\hat{\Psi}_{{\bf r}\alpha}^{\dagger}{h}_{0}\hat{\Psi}_{{\bf r}\alpha}\!+\hat{S}_{e}^{z}\int d{\bf r}g_{{\bf r}}^{z}\hat{S}_{{\bf r}}^{z}.\label{meanFH}
}
From the initial condition~\eqref{initial}, we can set the central spin to be a classical number $S_e^z=+1/2$. 
This approximation of neglecting the flip-flop interaction with setting $S_e^z=+1/2$ can be a useful starting point since the central spin is largely polarized during the dynamics in the case of a dense polarized spin bath \cite{BM07}. 
The resulting longitudinal Hamiltonian $\hat{H}_\parallel$ is quadratic in terms of the bosonic operators, where free bosons feel the mean-field potential $V_{\rm mean}=(V_T+V_S)/2$. Thus, the emergent Gaussian feature can be understood as a consequence of a large number of atoms stochastically occupying single-particle eigenstates of the mean-field Hamiltonian~$\hat{H}_\parallel$, which defines one of the key features of the Rydberg polaron~\cite{CF18,*RS18,KKS18}. We note that this feature is  determined entirely by the longitudinal couplings $g_{\bf r}^z$ and is thus not influenced by the anisotropy in the spin-interaction term. We also remark that the peaks in $A(\omega)$ in the limit of $\rho\to 0$ eventually match with bound-state energies for each of the Rydberg potentials, $V_T$ and $V_S$ \cite{BV09,GA14,RS16}.

The overlap $S(t)$ also accompanies oscillations, which can be best understood from the corresponding Fourier spectrum $A(\omega)$. 
 For a noninteracting setup described by a quadratic Hamiltonian (such as the triplet setting~\eqref{triplet} in Refs.~\cite{BV09,GA14,RS16,SM16,CF18,*RS18,KKS18}), the  frequency spacing of peaks matches the single-particle energy of the dominant bound state that has the largest overlap with the initial single-particle wavefunction. Correspondingly, the sharp peaks in the spectra indicate the formation of   Rydberg molecular states, where one, two or more environmental atoms are bound to and localized in the outermost well of the molecular potential.
 In contrast, in the present case of an interacting system, the central  spin interaction  causes the formation of a correlated many-body bound state. While the existence of sharp peaks in $A(\omega)$ still has its root in molecular physics, the simple explanation based on single-particle energies is no longer applicable to understand the underlying rich structures such as positions and spacings of the peaks and their sensitivity to the environmental density \cite{YA19L}. More specifically, the many-body feature manifests itself as the spin-interaction-induced renormalization of $A(\omega)$ from that of  noninteracting, bare molecular states  (see Fig.~\ref{fig_rev}). These results are qualitatively insensitive to specific details of the Rydberg potentials and thus should remain in a different choice of a Rydberg state as long as it has no angular momentum. 
 In particular, choosing a small principal quantum number $n$ for a Rydberg state, one can obtain a better resolution of the peaks in $A(\omega)$ whose energy scale can be an order of, e.g., MHz in contrast to a rather small energy scale in Fig.~\ref{fig_rev}. 
 
  \begin{figure}[t]
\includegraphics[width=56mm]{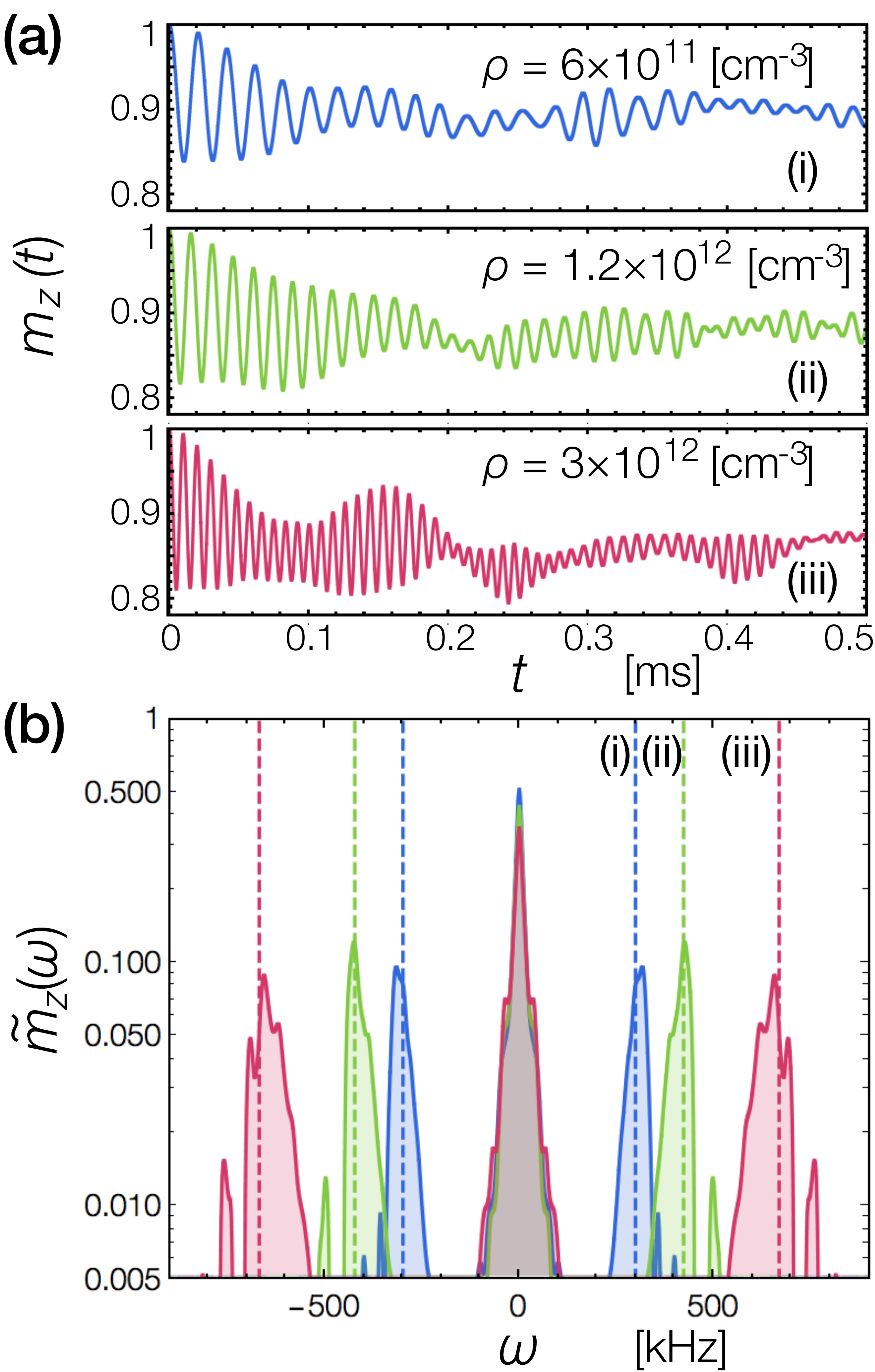} 
\caption{\label{fig2}
(a) The dynamics of the electron spin $m_z(t)=\langle\hat{\sigma}_e^z(t)\rangle$ of the Rydberg impurity interacting with environmental bosons via the anisotropic central  spin coupling. The results are obtained from the  variational approach and plotted at different densities $\rho$. (b) The corresponding Fourier spectra $\tilde{m}_{z}(\omega)$. The dashed lines indicate the values of the square root scaling $\omega\propto\sqrt{\rho}$. 
}
\end{figure}
 
In the noninteracting triplet setting, the anisotropy does not affect  values of  single-particle energies of the mean-field Hamiltonian~\eqref{meanFH} and thus the spectra remain the same as in the isotropic case \cite{YA19L}. In the present interacting problem, quantitative values of many-body bound-state energies can depend on the anisotropy since the interaction term includes the transverse  couplings $g_{\bf r}^{\perp}$, however, the qualitative features  remain the same.

Figure~\ref{fig2}(a) shows the corresponding spin dynamics $m_z(t)$ of the Rydberg electron at different densities $\rho$. 
Due to a large energy cost to flip the central spin coupled to a polarized environment,  only a small fraction of a many-body state with the opposite central spin $|\!\downarrow\rangle_e$ can be admixed. 
The resulting large polarization of the Rydberg spin again indicates that the flip-flop coupling does not play a significant role at the level of the mean-field feature of the absorption spectrum  (cf. Eq.~\eqref{meanF}). 

Remarkably, the spin dynamics is accompanied by fast, long-lasting oscillation that is reminiscent of the nondecaying oscillations found in the central spin problem \cite{BM07}. 
 To further study this feature, we plot the Fourier spectra $\tilde{m}_z(\omega)$ of the spin dynamics in Fig.~\ref{fig2}(b).
 The oscillation frequencies become higher as the density $\rho$ is increased and exhibit a square root scaling $\omega=a\sqrt{\rho}$ as indicated by the dashed lines in Fig.~\ref{fig2}(b). We find that the proportionality constant $a$ is larger than the corresponding value in the isotropic case \cite{YA19L} by the anisotropic factor $g_{\bf r}^{\perp}/g_{\bf r}^\parallel$ (which is chosen to be $\sqrt{2}$ in the present numerical calculation).  
This fact indicates that the frequency of the central  spin oscillation is mainly characterized by the transverse spin-exchange couplings $\hat{H}_{\perp}=(\hat{S}_e^{+}\int d{\bf r}g_{\bf r}^\perp\hat{S}_{\bf r}^-+{\rm H.c.})/2$. The square root scaling $\sqrt{\rho}$ can then be interpreted as a BEC enhancement factor that arises when the bosonic annihilation operator $\hat{b}_{\Downarrow}$ in $\hat{H}_\perp$ acts on a macroscopically occupied single-particle state of environmental bosons. 
Another interesting feature in Fig.~\ref{fig2}(a) is that the oscillations exhibit the collapse and revival in accordance with the motional wavepacket dynamics of bath atoms, whose time scale is roughly on the order of the binding energy. 
We note that a large scattering-length difference is crucial to observe these many-body features as it directly relates to the spin-exchange couplings $g_{\bf r}^{\perp}$.

Atoms coupled to the central spin via  strong flip-flop interactions are also subject to the strong orbital potential and thus typically ocuupy the dominant bound state localized in the outermost well of the Rydberg potential. 
We note that the observed square-root scaling with density is distinguished from the conventional linear scaling found in the standard central spin problem \cite{KA03,JS03,CWA04}, where the environmental spins are immobile and thus a macroscopically occupied state is absent. This weaker dependence of the oscillation frequency on a density is advantageous when one attempts to experimentally realize the predicted oscillation in atom clouds of inhomogeneous density. 
In practice, the magnetization dynamics might be measured by the Stern-Gerlach-like experiment. For instance, one may send back the Rydberg excitation to, e.g., the 6p state whose optical transition conserves the spin number, and then the transitioned atoms can be separated by using optical dipole traps. Measuring the population of the separated atoms with, for instance, using the ion microscopes, one could obtain information about the magnetization dynamics.

\begin{figure}[t]
\includegraphics[width=56mm]{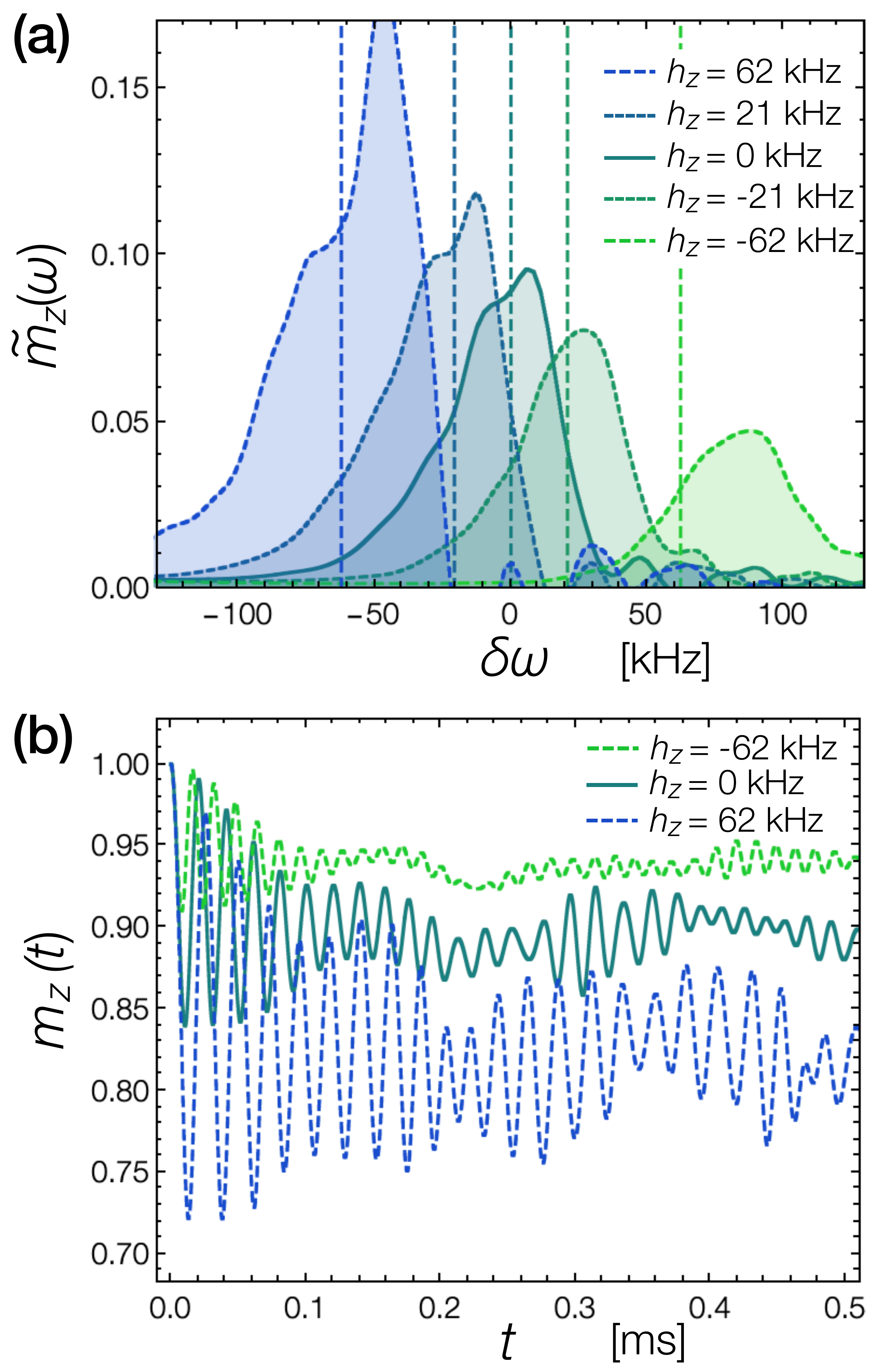} 
\caption{\label{fig3}
(a) The Fourier spectra $\tilde{m}_z(\omega)$ of the time evolution of the Rydberg electron spin with the anisotropic central  spin coupling at different magnetic fields $h_z$. The dashed lines indicate the linear scaling $\delta \omega=-h_z$. (b) The corresponding real-time dynamics of the Rydberg spin at different magnetic fields. In (a) and (b), we set $\rho=6\times 10^{11}$ cm$^{-3}$.
}
\end{figure}

The spin oscillation is strongly influenced by controlling an external magnetic field.
Figures~\ref{fig3}(a) and (b) show the spectra $\tilde{m}_{z}(\omega)$ and the real-time evolution $m_z(t)$ of the dynamics of the Rydberg electron spin at different magnetic fields $h_z$, respectively.   The oscillation frequencies of the spin oscillation can be controlled by changing $h_z$. We find that the shift $\delta\omega$ of the oscillation frequency from the zero-field value scales linearly with magnetic field, $\delta\omega= -h_z$, as indicated by the dashed lines in Fig.~\ref{fig3}(a). The amplitude of the oscillation is enhanced for a positive field $h_z>0$ for which the resonance is approached, while it is suppressed for the opposite sign of the field $h_z<0$.
These findings are consistent with the previously known magnetic-field dependences in the standard central spin problem \cite{BM07}. This fact shows that it is possible to manipulate the electron spin of dense Rydberg gases in the similar way as in solid-state qubits \cite{KA03,JS03,CWA04,HR08,WWM10,TAM12}. 
Furthermore, our findings also suggest that the Rydberg electron may be used to prepare and manipulate a mesoscopic, delocalized spin environment in a way analogous to localized nuclear-spin environments \cite{KFH05}.

\subsection{Monte Carlo results without orbital motion}
To elucidate the importance of taking into account the orbital motion of the environmental particles, we  analyze the case of  infinite-mass environmental particles  ($m\to\infty$). In this limit, the atomic orbital motion is completely frozen, and the position operators of the environmental bosons commute with the Hamiltonian. For each set of positions of environmental atoms $\{{\bf r}_i\}$, we can thus reduce the problem to the ordinary central spin problem:
\eqn{\label{infcsh}
\hat{H}_{m\to\infty}^{\left\{ {\bf r}_{i}\right\} }=V+\sum_{a=x,y}\hat{S}_{e}^{a}\sum_{i=1}^{N}g_{i}^{\perp}\hat{S}_{i}^{a}+\hat{S}_{e}^{z}\sum_{i=1}^{N}g_{i}^{\parallel}\hat{S}_{i}^{z},
}
where $V=\sum_{i=1}^{N}V_{0}({\bf r}_{i})$ and we denote the central spin couplings as $g_{i}^{\parallel}= g_{{\bf r}_{i}}^{z}$ and
$g_{i}^{\perp}= g_{{\bf r}_{i}}^{x,y}$.
Because of the conservation of the total spin-$z$ components, the time-evolved wavefunction is restricted to the sector of $\hat{\sigma}_e^z+2\sum_i\hat{S}_{i}^z=1-N$ and can  be parametrized as
\eqn{
|\Psi_{t}^{\{{\bf r}_{i}\}}\rangle&=&e^{-i\hat{H}_{m\to\infty}^{\left\{ {\bf r}_{i}\right\} }t/\hbar}|\Psi_{0}\rangle\nonumber\\
&=&\xi_{0}(t)|\uparrow\rangle_{e}|\Downarrow_{1}\cdots\Downarrow_{N}\rangle\nonumber\\
&+&\sum_{i=1}^{N}\xi_{i}(t)|\downarrow\rangle_{e}|\Downarrow_{1}\cdots\Uparrow_{i}\cdots\Downarrow_{N}\rangle.
}
The time-evolution equations of the variables $\xi$ are given by
\eqn{
i\hbar d_{t}\xi_{0} & =\left(V-\frac{G}{4}\right)\xi_{0}+\frac{1}{2}\sum_{i=1}^{N}g_{i}^{\perp}\xi_{i},\\
i\hbar d_{t}\xi_{i} & =\left(V+\frac{G-2g_{i}^{\parallel}}{4}\right)\xi_{i}+\frac{1}{2}g_{i}^{\perp}\xi_{0},
}
where $G=\sum_{i=1}^{N}g_{i}^{\parallel}$. This equation can 
be solved analytically by performing the Laplace transformation
\eqn{
\tilde{\xi}(s) & = & \int_{0}^{\infty}dt\xi(t)e^{-st},\\
d_{t}\xi & \to & s\tilde{\xi}(s)-\xi(0).
}
Using the initial condition $\xi_{0}(0)=1$ and $\;\xi_{i}(0)=0$ for $i=1,2,\ldots,N$, we  obtain the analytical expression for $\xi_0$, 
\eqn{\xi_{0}(t)\!=\!\frac{1}{2\pi i}\int_{{\cal C}}d\omega e^{-i\omega t-i\omega_{0}t}\left[\omega\!+\!\frac{G}{2}\!-\!\frac{1}{4}\sum_{i=1}^{N}\frac{g_{i}^{\perp2}}{\omega+\frac{g_{i}^{\parallel}}{2}}\right]^{-1},\nonumber\\
} 
where $\omega_{0}=(V+G/4)/\hbar$ and the contour ${\cal C}$
is chosen so that all the poles in the integral lie to its above; one can choose the contour that extends from $-\infty$ to $\infty$ along the real axis and is closed via a half circle in the upper complex plane.
After performing the integration, we find
\eqn{
\xi_{0}(t)=e^{-i\omega_{0}t}\sum_{l=1}^{N+1}w_{l}e^{-i\omega_{l}t}.
}
Here, $w_{l}=1/[1+\sum_{i=1}^{N}g_{i}^{\perp2}/(2\omega_{l}+g_{i}^{\parallel})^{2}]$
and $\left\{ \omega_{l}\right\} _{l=1}^{N+1}$ are the poles enclosed by the integration contour, which can be obtained from the  algebraic equation
\eqn{\label{betheroot}
\sum_{i=1}^{N}\frac{g_{i}^{\parallel}+(g_{i}^{\parallel2}-g_{i}^{\perp2})/(2\omega)}{2\omega_{l}+g_{i}^{\parallel}}=-1.
}
We note that Eq.~\eqref{betheroot} reproduces the previous result \cite{BM07} for obtaining the Bethe roots in the case of isotropic couplings $g_i^\perp=g_i^\parallel$. 
The overlap is  given by
\eqn{S^{\left\{ {\bf r}_{i}\right\} }(t)=\langle\Psi_{0}|e^{-i\hat{H}_{m\to\infty}^{\left\{ {\bf r}_{i}\right\} }t/\hbar}|\Psi_{0}\rangle=\xi_{0}(t),
}
leading to the absorption spectrum
\eqn{A^{\left\{ {\bf r}_{i}\right\} }(\omega)=\sum_{l=1}^{N+1}\delta(\omega-\omega_{0}-\omega_{l})w_{l}.\label{absinf}}
Similarly, the central  spin dynamics is given by
\eqn{m_{z}^{\{{\bf r}_{i}\}}(t)&\equiv&\langle\Psi_{t}^{\{{\bf r}_{i}\}}|\hat{\sigma}_{e}^{z}|\Psi_{t}^{\{{\bf r}_{i}\}}\rangle\nonumber\\
&=&|\xi_{0}(t)|^{2}-\sum_{i=1}^{N}|\xi_{i}(t)|^{2}.\label{mzinf}
}

To calculate the absorption spectrum appropriate for the  initial state~\eqref{initial}, 
we randomly generate sets of atomic positions \{${\bf r}_{i}$\}
according to the initial BEC wavefunction $\prod_{i=1}^{N}\psi_{{\rm ini},{\bf r}_{i}}$ and obtain the corresponding  $A^{\{{\bf r}_i\}}(\omega)$ from Eq.~\eqref{absinf}. This procedure is repeated to calculate  the ensemble average over different atomic configurations to yield
\eqn{A_{m\to\infty}(\omega)&=&\sum_{\{{\bf r}_{i}\}}{\rm Prob}[\{{\bf r}_{i}\}]A^{\{{\bf r}_{i}\}}(\omega).\label{absens}
}
Similarly, the  central  spin dynamics is obtained from  Eq.~\eqref{mzinf} and reads 
 \eqn{m_z^{m\to\infty}(t)&=&\sum_{\{{\bf r}_{i}\}}{\rm Prob}[\{{\bf r}_{i}\}]m_z^{\{{\bf r}_{i}\}}(t).\label{mzens}
}

\begin{figure}[t]
\includegraphics[width=56mm]{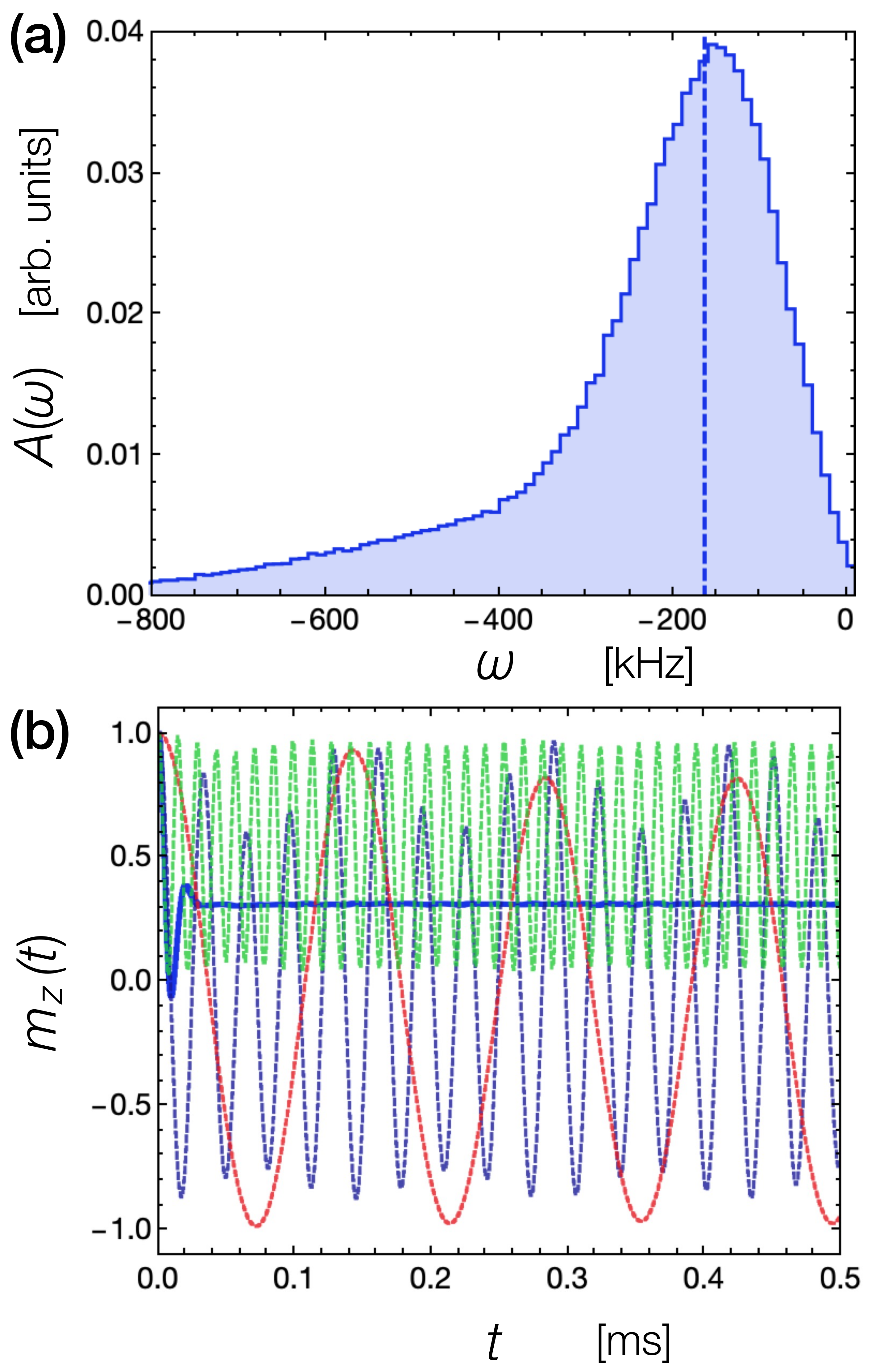} 
\caption{\label{fig4}
(a) Absorption spectrum $A(\omega)$ and (b) the corresponding central  spin dynamics $m_z(t)$ for environmental particles of infinite mass (i.e.,  localized environmental spins). The results are obtained from Eqs.~\eqref{absinf} and \eqref{mzinf} and by taking the ensemble average over $10^5$ different stochastic realizations of atomic configurations (cf. Eqs.~\eqref{absens} and \eqref{mzens}). In (a), the dashed line indicates the mean-field shift $\Delta_{\rm MF}$.
In (b), the blue thick solid curve indicates the ensemble-averaged result while the dashed thin curves show typical dynamics for single stochastic realizations. We set the density to $\rho=6\times 10^{12} {\rm cm}^{-3}$.
}
\end{figure}

Figures~\ref{fig4}(a) and (b) show the calculated  absorption spectrum $A(\omega)$ and the corresponding central  spin dynamics $m_z(t)$ for the infinite-mass environmental particles at typical parameters. The results are obtained by taking the ensemble average over $10^5$ different stochastic realizations of atomic configurations. As shown in Fig.~\ref{fig4}(a), the absorption spectrum exhibits the Gaussian feature which is the consequence of the average over various occupations of energy eigenstates of the stochastic central  spin Hamiltonian~\eqref{infcsh} (cf. Eqs.~\eqref{absinf} and \eqref{absens}). We note that, to obtain the spectrum, one still needs to solve the (integrable) many-body problem~\eqref{infcsh}, which should be distinguished from the single-particle calculations performed in, e.g., the triplet setting~\eqref{triplet}. To illustrate such a quantum aspect of the present problem, we plot typical single realizations of the central  spin dynamics with the fixed spin couplings (see the dashed thin curves in Fig.~\ref{fig4}(b)).  For each single realization, the spin exhibits nondecaying oscillations as previously discussed in the literature on the central spin coupled to polarized spin environments \cite{BM07}. However, after taking the ensemble average over different atomic configurations (cf. Eq.~\eqref{mzens}), these oscillations are averaged out, and the actual spin dynamics quickly relaxes to a stationary finite value as shown by the thick curve in Fig.~\ref{fig4}(b). 

We note that the key features found in the previous subsection, in which both the spin dynamics and the orbital motion are taken into account, are absent in the present  infinite-mass treatment. For instance, the peaks in the absorption spectrum do not appear as the Monte Carlo analysis cannot describe the formation of Rydberg molecular state. Also, the characteristic spin-precession dynamics is found to be absent in the infinite-mass treatment. These facts again highlight the importance of orbital motion of environmental particles to understand the dynamics of the RCSM.
 We remark that it would be useful to work out in a different single-particle basis such as position eigenstates for doing a direct comparison between the Monte Carlo analysis and the variational method in the large-mass regime. 

\section{Conclusions and outlook\label{sec4}}

We developed an efficient variational approach to solving quantum spin systems coupled to bosonic environments and applied this technique to analyze nonequilibrium properties of the Rydberg Central Spin Model (RCSM) proposed in the accompanying paper~\cite{YA19L}. 
The key element of this approach is the decoupling of the impurity spin using the canonical transformation~\eqref{unitary}. This transformation maps the conserved parity operator into one of the components of the impurity-spin operator and thus makes this spin component an integral of motion.  In the transformed frame, we then only need to consider the wavefunction of the bosons, which we approximate by the bosonic Gaussian state. This wavefunction includes all two particle correlations and requires the number of parameters growing only quadratically with the system size.
We utilized time-dependent variational principle to derive analytical expressions for the variational time-evolution equations (see Eqs.~\eqref{eqphi} and \eqref{eqGamma}  together with \eqref{Hphia} and \eqref{HGammaa}). These equations are general and can be readily applied to analyze the in- and out-of-equilibrium properties of a wide class of systems, in which quantum impurities are coupled to bosonic environments. 
As concrete examples, we applied our theory to analyze several dynamical aspects of the RCSM. In particular, we predicted sharp peaks in the absorption spectrum corresponding to dressed many-body molecular states, and long-lasting central  spin oscillations. To elucidate the crucial role of the orbital motion of atoms in these two features, we also analyzed the situation of immobile (i.e., infinitely heavy) bath atoms located at random positions. In this limit, the problem can be solved exactly for each stochastic realization; we then used the Monte Carlo sampling to take an ensemble average over possible atomic configurations. 
While the results are obtained by using the Rydberg potential of Rb atoms as an illustration, the qualitative physics found here will remain the same as long as the mean-field potential $V_{\rm mean}$ is attractive and supports a bound state, and also non-$s$-wave contributions do not play significant roles. Moreover, owing to the versatility of the developed variational approach, one can readily obtain quantitative predictions appropriate for concrete setups by using specific experimental parameters. 

Our study suggests several promising directions for future research. 
Firstly, in the context of solid-state systems, it is intriguing to apply the present variational approach to study the Bose Kondo problems \cite{FGM04,FS06,FFM11,FT15} in both in- and out-of-equilibrium regimes. In the field of ultracold atoms, our analysis can be extended to make predictions of site-resolved many-body dynamics as appropriate for experiments using  quantum gas microscopy~\cite{EM11,CM12,SP12,FuT15,KAM16,MM15,YA15,*YA16OL,*YA17multi,RY16,HS18}. The covariance matrix obtained in our approach can be expressed in the real-space basis, from which one can readily extract spatiotemporal dynamics of  correlation functions.
 It is intriguing to extend the present approach to analyze dissipative dynamics of quantum systems coupled to generic bosonic baths at finite temperatures. This can be done by generalizing the variational state to Gaussian density matrices. Another interesting direction is to extend our approach to Markovian open quantum systems subject to dissipation \cite{DAJ14,KEM12,BG13,YA18therm,*ZG18} and measurements \cite{HC93,LTE14,LR16,PYS15,YA17nc,*YA18fcs} by employing the variational principle appropriate for master-equation dynamics~\cite{CJ15,WH15}. Generalizing the decoupling canonical transformation, our approach can also be  applied to multiple impurities \cite{SP18} coupled to bosonic environments. In these ways, our approach can serve as a powerful theoretical tool to study a multitude of interesting quantum many-body problems.

 Secondly, it is intriguing to study the RCSM in a fermionic setup, where the interplay between the conventional central spin problem \cite{MG76,NVP00,TJM03,DJ04,EAY05,BM07,WZ07,CG07,LB08,BM10,WWM12,KEM12,FA13,RDA18} and fermionic Kondo physics \cite{HAC97,RL17,OK18} should offer another distinct class of  many-body  problems. Thirdly, our study suggests a promising direction of using Rydberg electrons to prepare and manipulate mesoscopic spin environments. While this technique has been implemented in solid-state devices \cite{KFH05} with a localized nuclear-spin environment, it remains an open question to what extent this method can be generalized to setups with delocalized (i.e., mobile) bath particles.
Finally, it merits further study to explore the impact of other experimental details arising from, e.g., nonzero angular momentum of the Rydberg electron and bath bosons or the hyperfine couplings.  These effects are expected to be important in several types of atoms \cite{ADA14,SH15,BF16,EF19}.
 We hope that our work stimulates further studies in these directions.

\begin{acknowledgements}
We are grateful to Shunsuke Furukawa, Tom Killian, Jesper Levinsen, Meera Parish, Masahito Ueda, and Shuhei Yoshida for fruitful discussions. Y.A. acknowledges support from the Japan Society for the Promotion of Science through Program for Leading Graduate Schools (ALPS) and Grant Nos.~JP16J03613 and JP19K23424, and Harvard University for hospitality. T.S. acknowledges the Thousand-Youth-Talent Program of China. 
R.S. is supported by the Deutsche Forschungsgemeinschaft (DFG, German Research Foundation) under Germany's Excellence Strategy -- EXC-2111 -- 390814868.
J.I.C. is supported by the ERC QENOCOBA under the EU Horizon 2020 program (grant agreement 742102).
E.D. acknowledges support from Harvard-MIT CUA, NSF Grant No. DMR-1308435, AFOSR Quantum Simulation MURI, AFOSR grant number FA9550-16-1-0323, the Humboldt Foundation, and the Max Planck Institute for Quantum Optics.
\end{acknowledgements}

\appendix
\section{Imaginary-time evolution\label{appimag}}
We here provide the variational equations for the imaginary-time evolution, which can be useful to analyze  ground-state properties. The exact form of the imaginary-time evolution in the transformed frame is given by
\eqn{|\Psi(\tau)\rangle=\frac{e^{-\hat{\tilde{H}}\tau}|\Psi(0)\rangle}{\left\Vert e^{-\hat{\tilde{H}}\tau}|\Psi(0)\rangle\right\Vert }.
}
Minimizing the error of the variational evolution from the exact one, we arrive at the following differential equation (cf. discussions above Eq.~\eqref{realtime}):
\eqn{d_{\tau}|\Psi_{{\rm GS}}(\tau)\rangle=-\hat{P}_{\partial}(\hat{\tilde{H}}-\langle\hat{\tilde{H}}\rangle_{{\rm GS}})|\Psi_{{\rm GS}}(\tau)\rangle.\label{imagpro}
}
Using the expressions~\eqref{GSderiv} and \eqref{HtildeGS} and comparing the linear and quadratic terms of $\hat{\boldsymbol \psi}$ on both sides of Eq.~\eqref{imagpro}, we  obtain
\eqn{d_{\tau}\boldsymbol{\phi}&=&-\Gamma{\cal H}_{\phi},\\
d_{\tau}\Gamma&=&\sigma^{{\rm T}}{\cal H}_{\Gamma}\sigma-\Gamma{\cal H}_{\Gamma}\Gamma.}
Integrating these variational equations together with the analytical expressions~\eqref{Hphia} and \eqref{HGammaa} of ${\cal H}_\phi$ and ${\cal H}_\Gamma$, one can analyze the ground-state properties in the limit of $\tau\to\infty$.

\section{Rydberg wavefunction\label{appryd}}
\begin{figure}[t]
\includegraphics[width=66mm]{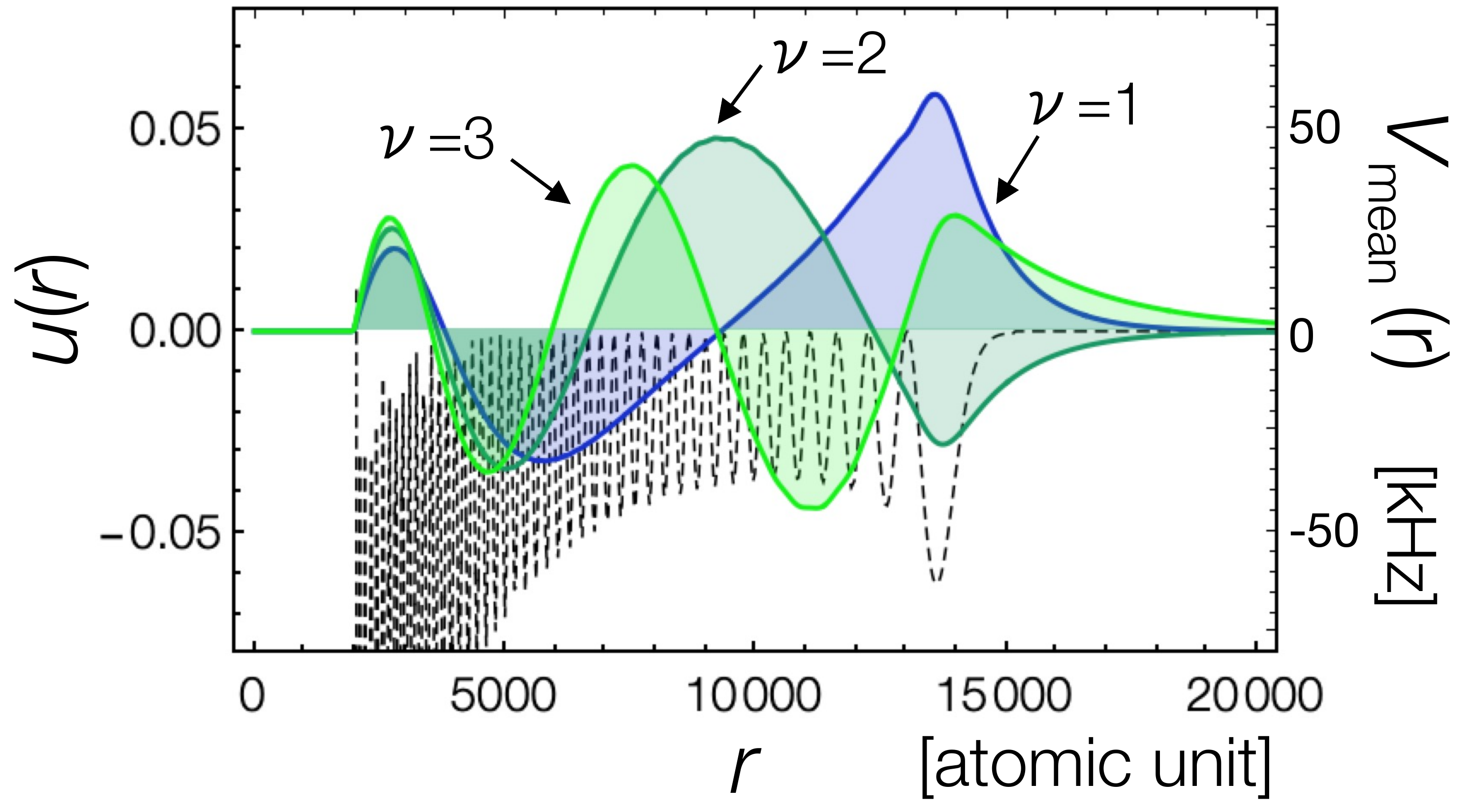} 
\caption{\label{appfig1}
Radial wavefunctions of the single-particle energy eigenstates for the mean-field Rydberg potential $V_{\rm mean}$.  Solution $\nu=1$ is the most dominant bound state having the largest overlap with the initial single-particle wavefunction. The black dashed curve indicates the spatial profile of the Rydberg potential. The bound-state energies for $\nu=1,2,3$ modes are $-21.5$, $-13.1$ and $-3.8$ (in the unit of kHz), respectively.
}
\end{figure}
We here briefly outline the calculation of single-particle energy eigenstates of atoms subject to a Rydberg potential.
We first note that, since the Hamiltonian is rotationally symmetric and the initial single-particle wavefunction $\psi_{{\rm ini},{\bf r}}$ has zero angular momentum, the system  resides in the sector with angular momentum $l=0$ during the course of the time evolution.  
The resulting radial single-particle Schr{\"o}dinger equation is thus given by
\eqn{\left[-\frac{\hbar^{2}}{2m}\frac{d^{2}}{dr^{2}}+V(r)\right]u_{n}(r)=\epsilon_{n}u_{n}(r),\label{radial}}
where $r$ denotes the radial coordinate, $V(r)$ is the Rydberg potential, and we introduce the radial wavefunction by 
$u_n(r)=\sqrt{4\pi r^{2}}\psi_{n{\bf r}}$. We numerically solve Eq.~\eqref{radial} in a spherical box of radius $R$ by imposing the boundary conditions $u_{n}(r_0)=u_{n}(R)=0$ with $r_0=2200$ and $R=10^5$ in the atomic units. We then obtain $N_b=80$  energy eigenstates from the lowest-energy solution. Figure~\ref{appfig1} shows three typical bound states for the mean-field pseudopotential $V_{\rm mean}=(V_T+V_S)/2$ for the scattering between the electron of the $^{87}$Rb$(87s)$ state and the surrounding ground-state $^{87}$Rb atoms. The state indicated by the label $\nu=1$ is localized around the outermost antinode of the potential and  corresponds to the dominant bound state having the largest overlap with the initial single-particle wavefunction. We note that the  contributions from the states with the principal numbers $n=1,2$, which are localized in the inner region and cutoff-dependent,  are tiny due to the vanishingly small overlaps with the initial state and thus are neglected in the analysis here.

The initial single-particle wavefunction is the lowest-energy state in the absence of an external potential and, by imposing the boundary conditions above, we obtain its solution as $u_{{\rm ini}}(r)=\sqrt{2/R}\sin(\pi r/R)$ where 
$u_{{\rm ini}}(r)=\sqrt{4\pi r^{2}}\psi_{{\rm ini},{\bf r}}$. We relate the total number $N$ of particles in the sphere with radial $R$ to the density $\rho$ by using the value of the initial wavefunction at the origin, i.e., 
$\rho=N|\psi_{{\rm ini},{\bf r}={\bf 0}}|^{2}.$
Finally, we note that the initial coefficients in Eq.~\eqref{initialphi} are obtained from
\eqn{\int d{\bf r}\psi_{n{\bf r}}^{*}\psi_{{\rm ini},{\bf r}}=\int_{0}^{R}dru_{n}^{*}(r)\sqrt{\frac{2}{R}}\sin\left(\frac{\pi}{R}r\right).
}

\section{Penalty term ensuring the spin conservation\label{apppenalty}}
In general, if the Hamiltonian $\hat{H}$ in the original frame satisfies a certain symmetry and has a conserved quantity $\hat{O}$, then the corresponding quantity $\hat{\tilde{O}}=\hat{U}^\dagger\hat{O}\hat{U}$ in the transformed frame is also guaranteed to be exactly conserved through the variational time-evolution equations. This  follows from the fact that the time-dependent variational principle ensures the symplecticity \cite{HJ11}. Yet, in practice, numerical errors accumulated during the integration of  highly nonlinear variational equations can  lead to an apparent violation of the conservation law.
To remedy this, it is useful to add a penalty term as a perturbation on the Hamiltonian so as to make sure that the conservation law is satisfied during the variational time evolution.  
In particular, in our problem, we ensure the spin conservation $\hat{\sigma}_{e}^{z}+\hat{\sigma}_{{\rm env}}^{z}+\hat{N}=\hat{\sigma}_{e}^{z}+2\hat{N}_{\Uparrow}=1$
by adding the  penalty term:
\eqn{\hat{V}=\lambda\left(\hat{\sigma}_{e}^{z}+2\hat{N}_{\Uparrow}-1\right)^{2}.
}
In the transformed frame, it is given by
\eqn{\hat{\tilde{V}}=\hat{U}^{\dagger}\hat{V}\hat{U}=\lambda\left(\hat{\sigma}_{e}^{x}\hat{{\rm P}}_{{\rm env}}+2\hat{N}_{\Uparrow}-1\right)^{2}.
}
Its expectation value with respect to the Gaussian state can be expressed as
\eqn{\langle\hat{\tilde{V}}\rangle_{{\rm GS}} 
 & =&\lambda\biggl[2-{\rm Tr}[P_{\uparrow}(\Gamma-{\rm I}_{4N_{b}})]-\boldsymbol{\phi}^{T}P_{\uparrow}\boldsymbol{\phi}\nonumber\\
 &+&\frac{1}{4}\left({\rm Tr}[P_{\uparrow}(\Gamma-{\rm I}_{4N_{b}})]+\boldsymbol{\phi}^{T}P_{\uparrow}\boldsymbol{\phi}\right)^{2}+\boldsymbol{\phi}^{T}P_{\uparrow}\Gamma P_{\uparrow}\boldsymbol{\phi}\nonumber\\
& +&\frac{1}{2}\left({\rm Tr}\left[P_{\uparrow}\Gamma P_{\uparrow}\Gamma\right]-{\rm Tr}\left[P_{\uparrow}\right]\right)+2\sigma_{e}^{x}{\cal P}\biggr],}
where $P_{\uparrow}={\rm I}_{2}\otimes((\sigma^{z}+1)/2\otimes{\rm I}_{N_{b}})$ and  ${\cal P}=2{\rm Tr}\left[P_{\uparrow}\Omega\right]-\langle\hat{{\rm P}}_{{\rm env}}\rangle_{{\rm GS}}$.
Its functional derivative with respect to the mean-field vector $\boldsymbol \phi$ is given by
\begin{widetext}
\eqn{{\cal H}_{\phi}^{V}&=&2\frac{\delta\langle\hat{\tilde{V}}\rangle_{{\rm GS}}}{\delta\boldsymbol{\phi}}=\lambda\left(2\left({\rm Tr}[P_{\uparrow}(\Gamma-{\rm I}_{4N_{b}})]+\boldsymbol{\phi}^{T}P_{\uparrow}\boldsymbol{\phi}-2\right)P_{\uparrow}\boldsymbol{\phi}+4P_{\uparrow}\Gamma P_{\uparrow}\boldsymbol{\phi}+4\sigma_{e}^{x}\frac{\delta{\cal P}}{\delta\boldsymbol{\phi}}\right),\label{penaltyphi}\\
\frac{\delta{\cal P}}{\delta\boldsymbol{\phi}} & =&-{\cal P}\Gamma_{B}^{-1}({\rm I}_{4N_{b}}+\Lambda)\boldsymbol{\phi}-4\langle\hat{{\rm P}}_{{\rm env}}\rangle_{{\rm GS}}\Gamma_{B}^{-1}{\cal S}\left[\Upsilon_{\uparrow}\right](\Gamma_{B}^{-1})^{{\rm T}}\boldsymbol{\phi},
}
\end{widetext}
where we introduce the matrix $\Upsilon_\uparrow$ as
\begin{widetext}
\eqn{\Upsilon_{\uparrow}=\left(\begin{array}{c}
{\rm I}_{2N_{b}}\\
-i{\rm I}_{2N_{b}}
\end{array}\right)\Sigma_{z}P_{\uparrow}({\rm I}_{2N_{b}},i{\rm I}_{2N_{b}}).}
\end{widetext}

Similarly, the functional derivative with respect to the covariance matrix $\Gamma$ can be calculated as
\begin{widetext}
\eqn{{\cal H}_{\Gamma}^{V}&=&4\frac{\delta\langle\hat{\tilde{V}}\rangle_{{\rm GS}}}{\delta\Gamma}=\lambda\left(2\left({\rm Tr}[P_{\uparrow}(\Gamma-{\rm I}_{4N_{b}})]+\boldsymbol{\phi}^{T}P_{\uparrow}\boldsymbol{\phi}-2\right)P_{\uparrow}+4P_{\uparrow}(\Gamma+\boldsymbol{\phi}\boldsymbol{\phi}^{T})P_{\uparrow}+8\sigma_{e}^{x}\frac{\delta{\cal P}}{\delta\Gamma}\right),\label{penaltygamma}\\
\frac{\delta{\cal P}}{\delta\Gamma} & =&{\cal S}\left[{\cal P}\Gamma_{B}^{-1}({\rm I}_{4N_{b}}+\Lambda)\left(\Phi-{\rm I}_{4N_{b}}\right)/2 +2\langle\hat{{\rm P}}_{{\rm env}}\rangle_{{\rm GS}}\Gamma_{B}^{-1}\Upsilon_{\uparrow}(\Gamma_{B}^{-1})^{{\rm T}}\left(2\Phi-{\rm I}_{4N_{b}}\right)\right].
}
\end{widetext}
The perturbations \eqref{penaltyphi} and \eqref{penaltygamma} are added to the functional derivatives~\eqref{Hphia} and \eqref{HGammaa}, respectively, and ensure the spin conservation during the variational calculations.

\bibliography{reference}
\end{document}